\documentclass[12pt]{article}

\usepackage{arxiv}
\usepackage[utf8]{inputenc} % allow utf-8 input
\usepackage[T1]{fontenc}    % use 8-bit T1 fonts
\usepackage{hyperref}       % hyperlinks
\usepackage{url}            % simple URL typesetting
\usepackage{booktabs}       % professional-quality tables
\usepackage{amsfonts}       % blackboard math symbols
\usepackage{amssymb}
\usepackage{bbm}
\usepackage{nicefrac}       % compact symbols for 1/2, etc.
\usepackage{microtype}      % microtypography
\usepackage{xcolor}         % colors
\usepackage{tcolorbox}
\usepackage{enumitem}
\usepackage{tikz}
\usepackage{amsmath}
\usetikzlibrary{decorations.pathmorphing}
\usetikzlibrary{fit}
\usetikzlibrary{shapes.arrows}
\usepackage{tcolorbox}
\usepackage{caption}
\usepackage{subcaption}
\usepackage{graphicx}
\usepackage{biblatex}
\usetikzlibrary{calc}
\usepackage{media9}
\begin{document}

\title{Dynamic Markov Blanket Detection for Macroscopic Physics Discovery}
\author{Jeff Beck \\ Noumenal Labs \\Department of Neurobiology, Duke University \And Maxwell J. D. Ramstead \\ Noumenal Labs\\ Queen Square Institute of Neurology, University College London, UK}

\maketitle

\begin{abstract}
The free energy principle (FEP), along with the associated constructs of Markov blankets and ontological potentials, have recently been presented as the core components of a generalized modeling method capable of mathematically describing arbitrary objects that persist in random dynamical systems; that is, a mathematical theory of ``every'' ``thing.'' Here, we leverage the FEP to develop a mathematical physics approach to the identification of objects, object types, and the macroscopic, object-type-specific rules that govern their behavior. To do so, we draw out the deep connections between Markov blankets, reinforcement learning, system identification theory, and macroscopic physics discovery. More specifically, we use the statistics of Markov blankets to operationalize two conditions for system equivalence in the literature, and to develop an approach to subsystem discovery, i.e., how to best partition a system into subsystems, and how to best classify (sub)systems. Using the statistics of Markov blankets, we demonstrate that two subsystems of a given system are weakly equivalent if their blankets share the same steady state statistics or reward rate; and strongly equivalent if the time evolution or paths of their boundaries have the same statistics. This allows us to formally define object types in terms of how they interact with their environment. It also allows us to reframe the problems of systems identification and macroscopic physics discovery as a problem of Markov blanket detection. We take a generative modeling approach and use variational Bayesian expectation maximization to develop a dynamic Markov blanket detection algorithm that is capable of identifying and classifying macroscopic objects, given partial observation of microscopic dynamics. This unsupervised algorithm uses Bayesian attention to explicitly label observable microscopic elements according to their current role in a given system, as either the internal or boundary elements of a given macroscopic object; and it identifies macroscopic physical laws that govern how the object interacts with its environment. Because these labels are dynamic or evolve over time, the algorithm is capable of identifying complex objects that travel through fixed media or exchange matter with their environment. This approach leads directly to a flexible class of structured, unsupervised algorithms that sensibly partition complex many-particle or many-component systems into collections of interacting macroscopic subsystems, namely, ``objects'' or ``things.'' We derive a few examples of this kind of macroscopic physics discovery algorithm and demonstrate its utility with simple numerical experiments, in which the algorithm correctly labels the components of Newton's cradle, a burning fuse, the Lorenz attractor, and a simulated cell.

\end{abstract}

%\tableofcontents % add TOC

\section*{Acknowledgments}
We thank Karl Friston for useful feedback. 

\newpage

\section{Introduction}

In this paper, we reconsider approaches to system identification and typification used in systems identification theory, engineering, and statistical physics using the tools of information theory, in particular, the constructs of Markov blankets and ontological potential functions that have been developed in the literature on the free energy principle (FEP). Inspired by recent derivations of classical and statistical mechanics from information theoretic principles \cite{niven2010, davis2015hamiltonian}, we present a novel derivation of the FEP, foregrounding how it can be used to write down ``ontological potential functions'' that define object types or phenotypes. This approach is based upon a consideration of the relative entropy formulation of Jaynes' principle of maximum caliber \cite{sakthivadivel2022geometry, ramstead2023bayesian, sakthivadivel2022worked}, with constraints imposed upon the boundary or Markov blanket of a macroscopic object or subsystem of a certain type. We show that these Markov blanket constraints fully characterize the interactions between an object or object type and the other objects in its environment, thereby formalizing the behavioral profile of a subsystem in terms of the effects that it has on other subsystems. This approach subsumes the more standard approaches to identification and typification of physical objects in contemporary reinforcement learning, system identification theory, and macroscopic physics discovery.

The problem of identification and typification of systems and subsystems is deceptively simple. In plain language, traditional systems identification theory asks the question: How can one characterize the complex interactions and behavior of a many-component system using a simple black box function approximator? In the standard approach to systems identification, the user identifies a subsystem as a connected subset of a large multi-component complex system and then characterizes the relationship between the inputs to and outputs from the subsystem. For example, in reinforcement learning, the inputs to a system are the observations of the agent and the outputs are its actions. The function that maps inputs to outputs is called the ``policy'' and in this context, two agents that execute the same policy perform the same actions given the same input conditions and, therefore, are considered to be equivalent. More generally, two subsystems with the same input-output relationship can then be said to be subsystems of the same type. 

Other than providing a means of typing subsystems, the utility of this approach lies in its ability to provide a compact description of a complex subsystem, by modeling it as a simple transfer function or as a low-dimensional dynamical system. This usually involves ``black-boxing'' the internal components of an overall system: that is, drawing a boundary around a many-object or many-component system, and then determining a (hopefully simplified) mathematical expression that summarizes the effect of the complex goings-on inside the boundary, in terms of their effects on the boundary of the subsystem, i.e., in terms of its input-output relationship to other subsystems. In this setting, the boundary itself is an arbitrary, user-defined boundary that separates the subsystem of interest from other subsystems. The definition of systems equivalence used in reinforcement learning is the one from systems identification theory: Two systems are then said to be of the same type if they have the same input-output relationship or policy, regardless of the details of their inner workings. 

In statistical physics, a similar notion of equivalence between macroscopic systems isused. Two systems are said to be of the same type if they can be modeled by the same dynamical system or energy function (Hamiltonian or Lagrangian). The main difference between the systems identification approach and the statistical physics approach is the latter's use of a coarse-graining procedure to simplify complex microscopic dynamics. This procedure begins by drawing an arbitrary boundary around a connected volume of microscopic particles and identifying macroscopic state variables that summarize the activity of the microscopic components (e.g., temperature, pressure, and density) inside the volume. Rules or laws that relate internal state variables to the net flux of conserved quantities at the boundary are then derived from knowledge of microscopic dynamics. Because flux is a conserved quantity, different volumes can then be connected to yield field equations that govern a larger system that is itself made up smaller subsystems that use the same flux variables. A macroscopic system of a given type can then be identified as consisting of the connected volumes that can be modeled using the same flux-state relationship, i.e., it is a bucket of water because each fluid element in the bucket has the same flux-state relationship. This extends the state space to a field that is dependent on the shape of the macroscopic object and establishes the bucket as providing boundary conditions. 

The free energy principle (FEP) has been proposed as a general mathematical modeling framework, unifying statistical mechanics and information theory, and providing a formal, biologically plausible approach to belief formation and updating as well as information processing. The FEP starts with a mathematical definition of a ``thing'' or ``object'': any object that when can sensibly label as such must be separated from its environment by a \textit{boundary}. Under the FEP, this boundary is formalized as a \textit{Markov blanket} that establishes conditional independence between that object and its environment. Within this framework, an object is defined not by physical states and fluxes or user labeled inputs and outputs, but rather by the flow of information across the Markov blanket. Strictly speaking, a Markov blanket defines the statistical boundary for a set of variables $Z\subset X$ as the minimal set $B\subset X$ such that $Z$ is conditionally independent of all variables not in $Z$ or $B$, given the blanket $B$. \cite{fristonparticular, ramstead2023bayesian}. In the FEP literature, Markov blankets are usually described as formalizing the notion of object, because the statistics of the Markov blanket fully characterize the input-output relationship between an object and other objects in its environment. 

This seemingly abstract definition of a boundary is, in fact, implicit in the definitions of a boundaries used in systems identification and statistical physics. In systems identification, the boundary of a system is directly defined by its inputs and outputs. Complete specification of the inputs and outputs of a system allows one to the treat the subsystem as a driving force on the system as a whole. Similarly, in statistical physics, knowledge of the fluxes into and out of a subsystem fully characterize both the subsystem and its impact on the rest of system, in the same way that initial conditions plus boundary conditions are sufficient to determine the evolution of the state variables. Thus, the notion that conditional independence is a property of boundaries between a subsystem and its environment is uncontroversial. In the FEP literature, it is therefore argued that we can define object types or phenotypes in terms of their Markov blankets. 

In the interest of precision, however, we note that it would be more accurate to say that the presence of a Markov blanket in a given system merely indicates a possible partitioning of the system into two interacting subsystems---and nothing else. Having identified a blanket, one can then make use of the fact that the blanket summarizes inputs and outputs of the associated subsystem to conclude that it is the statistics of the blanket that defines an object type. Recent work has made explicit the mathematical conditions under which the existence of a partition into subsystems with their Markov blankets can be guaranteed at steady state \cite{sakthivadivel2022weak, heins2022sparse}; and they are guaranteed to exist in the path-based formulation of the FEP \cite{sakthivadivel2022regarding, sakthivadivel2022weak}. Garnering significantly less interest is the question of how to discover these boundaries in the first place in a data-driven manner. Indeed, in most of the FEP literature, the focus is on explicating the dynamics of information flow in the presence of a blanket, and so the existence of a Markov blanket is usually \textit{assumed}, i.e., the existence and domain of the blanket are specified \textit{a priori}. When methods are proposed in the literature to identify Markov blankets, as they are, e.g., in \cite{friston2021parcels}, the focus is on an approximate blanket structure in the stationary distribution. This kind of blanket is rarely realized, even in systems that have actual blanket structure. Moreover, Markov blankets are always assumed to be static---or are associated with fixed components, giving rise to the false impression that things that display material turnover, such as traveling waves, flames, and living creatures cannot be modeled as having a Markov blanket \cite{raja2021markov, nave2025drive}. Taken together, all this suggests that the critical missing ingredient in the FEP literature is a procedure for identifying \textit{dynamic} Markov blanketed subsystems, which is thereby capable of describing a wide range of stationary and non-stationary phenomena, including flames, lightning bolts, organisms, and other systems that have transient or porous boundaries, and that can pop in and out of existence.

This requires the elucidation of a theoretical framework and associated class of inference algorithms that allow us to metaphorically ``carve the world at its joints'': that is, to partition complex many-component systems into macroscopic objects and object types, and discover the physical laws or macroscopic rules that govern the interactions between these objects. Ideally, such an unsupervised partitioning would (1) result in a compact, low-dimensional description of objects and their interaction, and (2) largely agree with human intuition, in that the partitions of systems into subsystems that it generates should largely be consistent with human intuitions about the associated perceptual phenomena. For reasons that will become apparent below, the class of algorithms that we are describing are \textit{dynamic Markov blanket detection algorithms}. The overall approach we take is based upon the formulation of the FEP in the space of paths or trajectories of systems over time, which is derived from Jaynes' principle of maximum caliber, coupled to the Markov blanket based definition of object and object type. This formulation leads directly to a notion of an ``ontological potential function,'' which is specified in terms of constraints on the blanket statistics and constraints on blanket dynamics, and which can be used as the basis for a taxonomy of object types. Here, ``blanket statistics'' refer to the typical summary of the dynamics of a subsystem in terms of its input-output relationship with its environment; while ``blanket dynamics'' refers to how the boundary itself changes over time. 

Starting from this definition of objects and object types, we consider a class of macroscopic generative models utilize two types of latent variables: (1) macroscopic latent variables that coarse-grain microscopic dynamics in a manner consistent with the imposition of Markov blanket structure, and (2) latent assignment variables that \textit{label} microscopic elements or observations in terms of their \textit{role} in a macroscopic object, its boundary, or the environment. Critically, these latent assignment variables are also allowed to evolve over time, in a manner consistent with Markov blanket structure. Finally, by taking a Bayesian approach to model discovery, we leverage to automatic Occam's razor effect of Bayesian inference to select the partitions of the system into subsystems, such that the global dynamics is as simple or low-dimensional as possible. 

In summary, we reformulate the problem of system identification as a Markov blanket detection problem. We take a generative modeling approach and use variational Bayesian expectation maximization to develop a dynamic Markov blanket detection algorithm that is capable of identifying and classifying macroscopic objects, given partial observation of microscopic dynamics. This unsupervised algorithm uses Bayesian attention to explicitly label microscopic elements according to their current role, as either the internal or boundary elements of a given macroscopic object; and it identifies macroscopic physical laws that govern how the object interacts with its environment. Because these labels are dynamic or evolve over time, the algorithm is capable of identifying complex objects that travel through fixed media or exchange matter with their environment. Crucially, this approach eliminates the need to impose arbitrary user-specified boundaries upon which systems identification typically relies, allowing for unsupervised segmentation of complex systems into collections of interacting macroscopic objects. Furthermore, by virtue of being based on discovering the statistics of Markov blankets, it automatically inherits the ability to identify object types, allowing us to classify subsystems in terms of the macroscopic rules or laws that govern how the object interacts with its environment. 

The rest of this paper is structured as follows. We first present an overview of Markov blankets and their use in the FEP. We then consider core elements of reinforcement learning, systems identification theory, and macroscopic physics discovery, mapping two notions of systems equivalence onto the statistics of Markov blankets, and discussing limitations. Following this, we return to Markov blankets under the FEP and discuss the mathematics of static and dynamic Markov blankets. We then present the Markov blanket detection algorithm and examine numerical work applying it to simple systems. We conclude with a discussion of implications of this work for the FEP broadly and directions for future work. We argue that the statistics of Markov blankets as formulated in the FEP literature provide us with the mathematical apparatus needed to establish the notion of an ``ontological potential function,'' i.e., a function that rigorously defines object types via boundary constraints.

\section{The free energy principle: Core elements}

\subsection{Markov blankets in the formulation of the free energy principle}

At a high level, the standard formulation of the FEP starts from the equations of statistical physics, with \textit{Markov blanket} structure imposed. The notion of a Markov blanket was originally introduced by Pearl \cite{pearl1998} as a means of identifying the complete set of random variables that impact inference regarding the value of a given set of ``internal'' random variables.\footnote{Note that here we are using Markov blanket and Markov boundary interchangeably. In some parts of the literature, a Markov \textit{blanket} refers to any set that establishes the desired conditional independence structure, while the Markov \textit{boundary} is the minimal set. \cite{pearl1998}} Pragmatically, knowledge of a Markov blanket for each node in a graphical model can be used to identify the structure of message passing algorithms used for efficient probabilistic inference. The property is inherited directly from the definition of a Markov blanket of set of ``internal'' random variables $z \subset X$ as set $b \subset X$ such that $z \perp s | b$, where $s$ is the complement of the union of $z$ and $b$, or equivalently:

\begin{equation}
    p(s,z|b) = p(s|b)p(z|b)
\end{equation}

\noindent In a directed graphical model, the Markov blanket of a set of nodes $Z$ consists of all the parents of the nodes in $Z$, all the children of nodes in $Z$, and all the parents of the children of the nodes in $Z$. This establishes a conditional independence relationship between nodes in $Z$ and all of the other nodes not in the blanket of $B$, i.e. $S = (Z\cup B)^c$, where the superscript $c$ denotes the complement, 

\begin{eqnarray}
p(z|b) & = & p(z|s,b) \\
p(s|b) & = & p(s|z,b)
\end{eqnarray}

\noindent where the lowercase refers to the values of the random variables in the corresponding set.

In the FEP, this Markov blanket structure is realized by assuming that the dynamics of a microscopic system can be partitioned into three subsets of variables: internal variables ($z$), boundary variables ($b$), and external or environmental variables ($s$), such that:

\begin{align}
    \frac{ds}{dt} &=  f_s(s,b) + \eta_s \\
    \frac{db}{dt} &=  f_b(s,b,z) + \eta_b \\
    \frac{dz}{dt} &=  f_z(b,z) + \eta_z
\label{eq:sbz}
\end{align}
    
\noindent where the $\eta$'s indicate noise that is independent across the $s,b,z$ variables. This results in the desired posterior probability distribution over trajectories or paths (indicated by subscript $\tau$):

\begin{equation}
p(s_\tau,z_\tau|b_\tau)p(b_\tau) = p(s_\tau|b_\tau)p(z_\tau|b_\tau)p(b_\tau)
\end{equation}

Conditional independence follows from the absence of any direct causal interaction between $s$ and $z$, but can be intuitively understood as resulting from that fact that, if the path $b_\tau$ is observed, then it can be treated as a known driving force to two independent subsystems. It is important to note that the Markov blanket associated with this system applies to paths (i.e., to the time evolution or trajectory of a system), and not to steady state distributions. Also important to note is that dynamical systems that have blanket structure (i.e., that conform to Eq. \ref{eq:sbz}) do not generally result in steady state distributions that also have blanket structure. See \cite{fristonparticular} for some exceptions to this general rule. 

The link between the Markov blanket and subsystem or object type also follows directly from the conditional independence relationship. This is because two objects whose boundaries follow the same path must, by definition, have the same effect on the environment, regardless of the details of their internal dynamics. As a result, one can define an object type by the path statistics of a Markov blanket. See also \cite{sakthivadivel2022worked}. Crucially, we note that this statistical definition of object type is consistent with the definition of object type used in systems identification theory, which is standard in reinforcement learning. This is because the blanket statistics fully characterize the interactions between a subsystem and its environment, and thus, include both the inputs and outputs of the subsystem. As such, two objects with very different internal structure and states, but with the same boundary statistics, will also have the same input-output relationship and will interact with their environment in precisely the same way. This is trivially observed in the simple case where one can partition the blanket into active states that directly influence the external variables and sensory states that directly influence the internal variables, i.e. $b = \left\{a,o\right\}$. In this case, straightforward application of Bayes rule to the Markov blanket based definition of object type, $p(b_\tau)=p(a_\tau, o_\tau)$, allows for the direct computation of $p(a^t| o_{\tau<t})$, which we recognize as the agent's policy or the subsystem's response function. 

It is worth noting that this Markov-blanket-statistics-based formulation of systems equivalence offers a more complete description than the one from systems identification, in that it contains more than just the policy of the agent or object in question. This is because the equations are symmetric and so the blanket statistics also encode the ``policy'' of the external system, $p(o^t| a_{\tau<t})$. This makes it clear that the Markov-blanket-based definition of an object type is environment-specific. 

\begin{figure}[ht]
  \centering
  \begin{subfigure}{0.4\textwidth}
    \centering
    \begin{tikzpicture}[scale=1]
      % red particles
      \foreach \x in {1,...,12}
          \draw[fill=red, radius=0.2] ({\x*30}:1.8) circle;
      % blue particles
      \foreach \x in {1,...,3}
          \draw[fill=blue] ({\x*120}:0.6) circle (0.2);
      % Green particles
      \foreach \x in {1,...,6}
          \draw[fill=green] ({\x*60}:1.2) circle (0.2);
      % Ring connections
      \draw[thin] (60:1.2) -- (120:1.2) -- (180:1.2) -- (240:1.2) -- (300:1.2) -- (360:1.2) -- cycle;
      \foreach \x in {1,...,12}
          \draw[thin] ({30*(\x)}:1.8) -- ({30*(\x)+30}:1.8);
      \foreach \x in {1,...,6}
          \draw[thin] ({(\x)*60}:1.2) -- ({\x*60}:1.8); 
      \foreach \x in {1,...,6}
          \draw[thin] ({(\x)*60}:1.2) -- ({\x*60+30}:1.8); 
      \foreach \x in {1,...,6}
          \draw[thin] ({(\x)*60}:1.2) -- ({\x*60-30}:1.8); 
      % blue to green connections
      \foreach \x in {1,...,3}
          \draw[thin] ({\x*120}:0.6) -- ({(\x)*120-60}:1.2);
      \foreach \x in {1,...,3}
          \draw[thin] ({\x*120}:0.6) -- ({(\x)*120+60}:1.2);
      \foreach \x in {1,...,3}
          \draw[thin] ({\x*120}:0.6) -- ({(\x)*120}:1.2);
      % blue to blue connections
      \draw[thin] (120:0.6) -- (240:0.6);
      \draw[thin] (240:0.6) -- (360:0.6);
      \draw[thin] (360:0.6) -- (120:0.6);
    \end{tikzpicture}
    \caption{'Actual'}
    \label{fig:actual}
  \end{subfigure}
  \hfill
  \begin{subfigure}{0.4\textwidth}
    \centering
    \begin{tikzpicture}[scale=1]
      % red particles
      \foreach \x in {1,...,12}
          \draw[fill=red, radius=0.2] ({\x*30}:1.8) circle;
      % blue particle
      \draw[fill=blue] (0, 0) circle (0.4);
      % Green particles
      \foreach \x in {1,...,6}
          \draw[fill=green] ({\x*60}:1.2) circle (0.2);
      % Ring connections
      \draw[thin] (60:1.2) -- (120:1.2) -- (180:1.2) -- (240:1.2) -- (300:1.2) -- (360:1.2) -- cycle;
      \foreach \x in {1,...,12}
          \draw[thin] ({30*(\x)}:1.8) -- ({30*(\x)+30}:1.8);
      \foreach \x in {1,...,6}
          \draw[thin] ({(\x)*60}:1.2) -- ({\x*60}:1.8); 
      \foreach \x in {1,...,6}
          \draw[thin] ({(\x)*60}:1.2) -- ({\x*60+30}:1.8); 
      \foreach \x in {1,...,6}
          \draw[thin] ({(\x)*60}:1.2) -- ({\x*60-30}:1.8); 
      % blue to green connections
      \foreach \x in {1,...,6}
          \draw[thin] (0, 0) -- ({\x*60}:1.2);
    \end{tikzpicture}
    \caption{Equivalent}
    \label{fig:equivalent}
  \end{subfigure}
  \caption{The Markov blanket definition of object equivalence. The relationship between the boundary (green) and environment (red) are fixed. The objects are equivalent if replacing the internal variables (blue) and their connections to the boundary with some other set of variables and connections without affecting the distribution of boundary paths, $p(b_\tau)$.}
  \label{fig:mbequiv}
\end{figure}
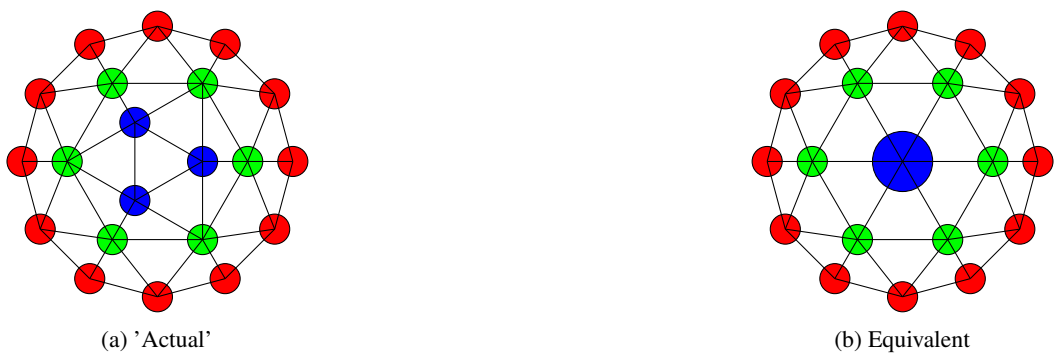

\subsection{The maximum caliber route to an ontological potential function}

The utility of any definition lies in its predictive power. Here, we show that, when combined with Jaynes' principle of maximum caliber \cite{jaynes1980minimum, davis2015hamiltonian, niven2010}, this blanket statistics based definition of an object leads directly to an ``ontological potential function'' that formalizes the notion object type in terms of the macroscopic rules of behavior necessary to instantiate an object of a given type, where type is defined by blanket statistics \cite{sakthivadivel2022geometry, ramstead2023bayesian}. That is, given path statistics on a Markov blanket, $p(b_tau)$, Jaynes' principle can be used to identify an energy function and associated Lagrangian that specify the dynamics of the environment, boundary, and object variables that give rise to an object of that type. This is consistent with the notion of system typing used in physics, wherein physical systems are defined by the energy functions or stationary actions that they minimize. (For a similar argument based on the maximum entropy principle, see \cite{kiefer2020psychophysical}.) Here, we show that boundary statistics lead directly to an object-specific energy function, an ontological potential function that corresponds to the generalized free energy and associated Lagrangian. This closes the gap between the Markov blanket based definition of an object type and the notion of type used in statistical physics and leads naturally to a \textit{typology of systems} based on statistical constraints imposed on their Markov blankets \cite{ramstead2024approach}. 

The objective function that one obtains from combining blanket statistics with Jaynes' principle of maximum caliber is precisely the \textit{free energy}, providing the basis for an alternative derivation of the FEP. To this end we take inspiration from recent derivations of statistics physics from purely information theoretic considerations \cite{niven2010,davis2015hamiltonian}, and impose boundary constraints within the \textit{maximum caliber} framework. We show that this ultimately results in a free energy minimization problem, consistent with the core elements of the path based formulation of the FEP \cite{ramstead2023bayesian}. Jaynes' principle of maximum caliber is an information theoretic formulation that extends the constrained maximum entropy approach \cite{jaynes1957} to the space of paths or trajectories through state space. This information theoretic approach is commonly used in statistical inference, to determine the most parsimonious model of some data; where we maximize entropy or caliber under constraint derived from data or known physical laws \cite{jaynes1980minimum}. More recently, Jaynes' principle applied to paths through state space has been shown to provide an information theoretic foundation upon which one can derive the equations upon which physics is based; that is, given an appropriately chosen set of constraints, optimizing Jaynes' caliber objective leads directly to all the core equations of classical, statistical, and quantum physics in such a way that the deep relationships between action, energy, work, and heat (Hamiltonian or Lagrangian functions) are made plain \cite{davis2015hamiltonian, niven2010}. It can also be used to directly derive powerful theorems in non-equilibrium statistical physics, including Crook's theorem, Noether's theorem, the Jarzynski inequality, and the second law of thermodynamics itself \cite{niven2010, jarzynski2012equalities, gonzalez2016}.

The near uniform applicability of Jaynes' maximum caliber principle, when combined with the blanket-statistics-based definition of an object type allows us to incorporate non-stationary and non-ergodic systems into the FEP modeling framework, something that had seemed difficult to model using previous formulations (see \cite{bruineberg2022emperors, raja2021markov, dipaolo2022forking, nave2025drive}). As we will see, this flexibility will allow us to define---and ultimately discover, in a data-driven manner---the dynamic or wandering Markov blankets that are required to model non-stationary, changing, and mobile things such as flames, lightning bolts, and traveling waves, as well as objects that exchange matter with their environments. 

\subsubsection{Jaynes' principle and ontological potential functions}

Mathematically, maximum caliber begins with a probabilistic, and often coarse, characterization of the laws of physics, $p(\dot{x},x,t)$. It then supposes that we have additional information about how a particular system works, which is given by certain constraints that take the form of time dependent expectations 

\begin{equation}
F(t) = \left<f(\dot{x},x,t)\right>
\end{equation}

These expectations provide ``instantaneous'' constraints and, when combined with prior knowledge of dynamics, effectively define a system type. For example, stationary geometric constraints $<f(x)> = C, \forall t$ lead to the traditional potential function of classical mechanics, while the kinetic term arises from Newton's laws, as expressed via $p(\dot{x},x,t)$. A probability law $q(\dot{x},x,t)$ associated with that system type is then obtained by maximizing the relative path entropy, with the constraints enforced by Lagrange multipliers, $\lambda(t)$, i.e. 

\begin{equation}
S[q(\cdot),\lambda] = -KL(q(\dot{x},x,t),p(\dot{x},x,t)) - \left< \int \lambda(t) \cdot  f(\dot{x},x,t) dt\right>_{q(\cdot)}
\label{eq:maxent}
\end{equation}

\noindent along with the unstated additional constraint that $q(\cdot)$ is a well-defined probability distribution, i.e. the integral over all paths is 1. Written in terms of the Lagrange multipliers, optimization of $S[\cdot]$ results in 

\begin{equation}
q(\dot{x},x,t) = \frac{1}{Z[\lambda(t)]}p(\dot{x},x,t)\exp\left( -\int \lambda(t) \cdot f(\dot{x},x,t)\right)
\end{equation}

where the path entropy at the maxima is given by 

\begin{eqnarray}
S_{max} & = & \log Z[\lambda(t)] + \int \lambda(t)\cdot F(t)dt \\
\mathrm{Entropy} & = & - \mathrm{Free \hspace{3pt} energy} + \mathrm{Energy} 
\end{eqnarray}

with associated dimensionless Lagrangian and action:

\begin{align}
L(\dot{x},x,t) = \log p(\dot{x},x,t) + \lambda(t) f(\dot{x},x,t) \\
A\left[x_\tau\right] = \int dt L(x,\dot{x};t)
\end{align}

This relationship between expected energy, entropy, and free energy allows us to cast the maximization of caliber as a \textit{free energy minimization}, consistent with recent formulations of the FEP \cite{friston2023path}. Moreover, armed with the Lagrangian, one directly obtains the associated Hamiltonian with the interpretation that Hamiltonian dynamics yield the most probable path \cite{davis2015hamiltonian}. Constraints that are also independent of time coupled with the assumption of time translation symmetry leads directly to constant $\lambda$ and the identification of a conserved energy ($\lambda\cdot F$). 

Critically, we can interpret this log partition function form of free energy as a \textit{potential function}, the partial derivatives of which result in generalized notions of \textit{heat and work} \cite{niven2010}. In this maximum caliber framework, it is the constraints that ultimately lead to this potential function, as well as that Langevin and Hamiltonian dynamics that define physical systems. This implies that the constraints themselves can be thought of as defining system type. For this reason, we conclude that this free energy functional \emph{is} an \textit{ontological potential function} for noisy dynamical systems, and a central component of a generic definition of ``every'' ``thing.''

\subsubsection{The Markov blankets of ``every'' ``thing''}

What the FEP approach adds to Jaynes' principle of maximum caliber is a physical definition of the boundary of ``every'' ``thing.'' That is, the FEP adds to maximum caliber a way of representing the notion that boundaries are not fictive (e.g., mere flux relations between volumes), but rather, that they correspond to distinct, seperable objects. This is inherited from the notion that subsystems are defined by \textit{boundary or blanket constraints}.  

Combining the notion of an ontological potential function, as defined above, with the Markov blanket definition of a subsystem type leads directly to a simple definition of a subsystem or ``object'' or ``thing'' in a given environment:

\begin{tcolorbox}[
  colback=white,              % Background color
  colframe=black,             % Frame color
  fonttitle=\bfseries,        % Title font
  title={Maximum caliber definition of an object},    % Title text
]

\begin{enumerate}
  \item A time dependent set or manifold $B \subset X \otimes T$, parameterized by $\Omega_B(t) \subset X$ that maintains Markov blanket structure with respect to the prior on unconstrained dynamics $p(x,\dot{x})$
  \item A set of instantaneous constraints applied to both the set $\Omega_B(t)$ and the elements of $x\in \Omega_B(t)$ 
\end{enumerate}
\end{tcolorbox}

It is easy to show that imposing blanket constraints does not disrupt the conditional independence structure inherited from the prior dynamics. While it is possible to enforce the constraint that a particular $p(b_\tau)$ be realized directly\cite{davis2015hamiltonian}, we find it more intuitive to re-represent the blanket distribution by a (possibly infinite) set of instantaneous constraints on expectations. Of course, the constraints needed to represent an arbitrary $p(b_\tau)$ are not necessarily the kind of ``instantaneous'' constraints that are typically used with the maximum caliber objective. Instantaneous constraints, however, are preferred because they lead to causal dynamics and are therefore considered ``physically realizable.'' We restrict ourselves to this class of instantaneous constraints by defining a time-dependent blanket, $\Omega_B(t)\subset X$, and imposing constraints of the form

\begin{equation}
    F_B(t) = \left< f_B(\dot{x},x,t,\Omega_B(t)|_{x\in\Omega_B(t)} \right>
\end{equation}

Note that when applied to a blanket that is a manifold, geometric constraints simply add an additional flux term to the associated Langevin equations, but do not otherwise change their structure. This arises from the time derivative of the indicator function that restricts boundary constraints to elements in the boundary. This is true regardless of whether or not the blanket is connected or persists for any length of time. Note also that judicious choice of constraint functions, $f_B(\cdot)$, can be used to place constraints on the shape and evolution of the manifold $\Omega_B(t)$. The flexibility of this approach follows from the fact that the boundary $\Omega_B(t)$ can be either specified or treated as a random variable with support on the set of Markov blankets specified by the prior $p(\cdot)$. 

Before moving on, we note some connections to systems identification and reinforcement learning. As previously shown, equivalence of Markov blanket statistics implies policy equivalence. In reinforcement learning, this is often referred to as strong equivalence between two agents. Since policies can be derived from the blanket statistics, equivalence of Markov blanket statistics implies policy equivalence; and therefore, agents that share Markov blanket statistics are agents of the same type. Reinforcement learning also has a notion of weak equivalence that is associated with agents that achieve the same reward rate. In an infinite horizon setting, with a reward function that only depends on actions and outcomes, agents that are defined by stationary boundary statistics necessarily achieve the same reward rate. The converse is also true: namely, agents with the same stationary boundary statistics have the same reward function, under the assumptions imposed by the maximum entropy inverse reinforcement learning paradigm \cite{ziebart2008maximum, snoswell2021revisiting}. This means that weakly equivalent agents are associated with constraints on the stationary distribution of the boundary, $\tilde{p}(b)$. A more direct link to reinforcement learning can also be seen by noting the link between Jaynes' maximum caliber objective and that of KL control theory \cite{kappen2012optimal, Todorov2010InverseOC}. This makes it plain that the Markov blanket statistics based notion of object type and associated ontological potential function subsumes the relevant notions of systems equivalence used in both systems identification and contemporary reinforcement learning. 

In short, this maximum caliber, blanket-statistics-based definition of object type has the desired property that it is widely applicable, consistent with systems identification theory, and associated with a consistent objective function, namely, the free energy. While the specific details of the constraints imposed by $F_B$ indicate an object's particular type, gross properties of the constraint functions themselves can be used to classify different kinds of objects. It also suggests the possibility of developing a taxonomy of different object types based on the domain of the blanket and the kinds of ``ontological'' or object-type-specific constraints imposed. For example, it is standard in the statistical physics literature to refer to constraints that only depend upon $x$ as geometric constraints. The approach outlined here elucidates at least three degrees of freedom along which constraints can be defined: the shape or topology of the boundary $\Omega_B(t)$, the dynamics of the boundary, and the statistics of the boundary. For example, the boundary of a cell is, topologically speaking, a spherical surface that is dynamic because it exchanges matter with its environment, and that has stationary boundary statistics to the extent that achieves homeostasis in its environment. A rigid body, on the other hand, has no internal states \textit{per se}, and thus has a spherical ``boundary'' that is static in the sense that it does not change shape, but dynamic in the sense that it can move.  

\subsection{A simple example: the flame}

A common criticism of the Markov blanket definition of a ``thing'' has been that this definition does not apply to flames \cite{nave2025drive, raja2021markov}. This misconception results from the false assumption that Markov blankets must be static or tied to matter. However, the formalism presented here neatly captures flames and other traveling waves, due to the flexibility endowed via a consideration of dynamic boundaries. 

For example, consider a simple flame burning down a one dimensional fuse as an unsteady traveling wave (we provide a numerical example in the following sections). Here, we will define the boundary to be the point that separates burned from unburned regions, $y_b(t)$. The constraint we will impose is that the temperature at this point should correspond to the ignition temperature of the exothermic chemical reaction that drives the flame

\begin{equation}
    \theta_{ig} = \left< \int dy' \delta(y'-y_b(t))T(y',t) \right> 
\end{equation}

Implementing this constraint results in a Lagrangian and maximum a posteriori (MAP) estimate of temperature given by:

\begin{align}    
    \mathcal{L}(T,\dot{T},\nabla^2 T,y_b(t),t) &= \log p(\cdot) + \lambda(t)\int dy'\delta(y'-y_b(t)) T(y',t) \\
    \left(\frac{\partial}{\partial t} - \frac{\partial^2}{\partial y^2} + h\right) T_{map} &= \sigma_p^2\lambda(t)\delta(y-y_b(t))
\end{align}

\noindent Thus, the imposition of the constraint results in a point heat source at the boundary, with magnitude proportional to the Lagrange multiplier that implements the constraint. Here, $\sigma_p^2$ represents the variance of the deviation from a prior that favors the heat equation for a 1 dimensional solid subject to Newton's law of cooling. Recall that energy is given by the constraint: $\int dt \lambda(t)\theta_{ig}$ allowing for the interpretation of $\lambda(t)$ as the energy that must be injected into the system in order to keep the flame moving at a specified velocity $\dot{y}_f(t)$.

While this example illustrates how constraints and boundaries can work together to lead to consistent and sensible dynamics, it is not quite what we would like. Ideally, we would like a probability distribution over paths that includes a distribution over the boundary location and state. This requires a prior that operates on the location of the boundary, and that in general can include additional constraints. This allows us to treat $y_b(t)$ in the same way as $T(y,t)$, resulting in an expansion of the set of equations that must be simultaneously solved in order to determine MAP dynamics for both temperature and flame speed. For example, if we assume \textit{a priori} that the flame speed is normally distributed, then we acquire an additional Euler-Lagrange equation that relates flame speed fluctuations to heat release and temperature flux at the boundary:

\begin{equation}
\frac{d^2}{dt^2} y_b(t) = \sigma^2_{y_b}\lambda(t)\frac{\partial}{\partial y} T_{map}(y_b(t),t).
\end{equation}
 
\subsection{Summary so far}

In practice, how do you use an ontological potential function? We begin by specifying a physics via a prior on dynamics $p(\dot{x},x,t)$, and we then specify the statistical properties of a boundary via constraints. The dynamical system that results from this specification is guaranteed to give rise to the kind of object specified by the boundary statistics. This can be interpreted as answering the questions: (1) how must microscopic elements (both internal and external to the subsystem) organize themselves, or (2) how must energy be injected into and dissipated by the system, in order to instantiate a boundary of the specified type.\footnote{Of course, the dynamics obtained thereby are not unique: There might indeed be lower entropy solutions. However, the one obtained via this maximum caliber approach can be guaranteed to be the most general.} 

However, this method of instantiating the desired object still does not allow us to determine which particular subsets of a given dynamical system system \emph{should} qualify as an object. It only tells us something about systems that support particular kinds of objects that we have already identified. Indeed, despite their utility, all the methods just reviewed do not adequately address the problem of subsystem identification and typification, precisely because the boundaries are effectively \textit{user defined}.

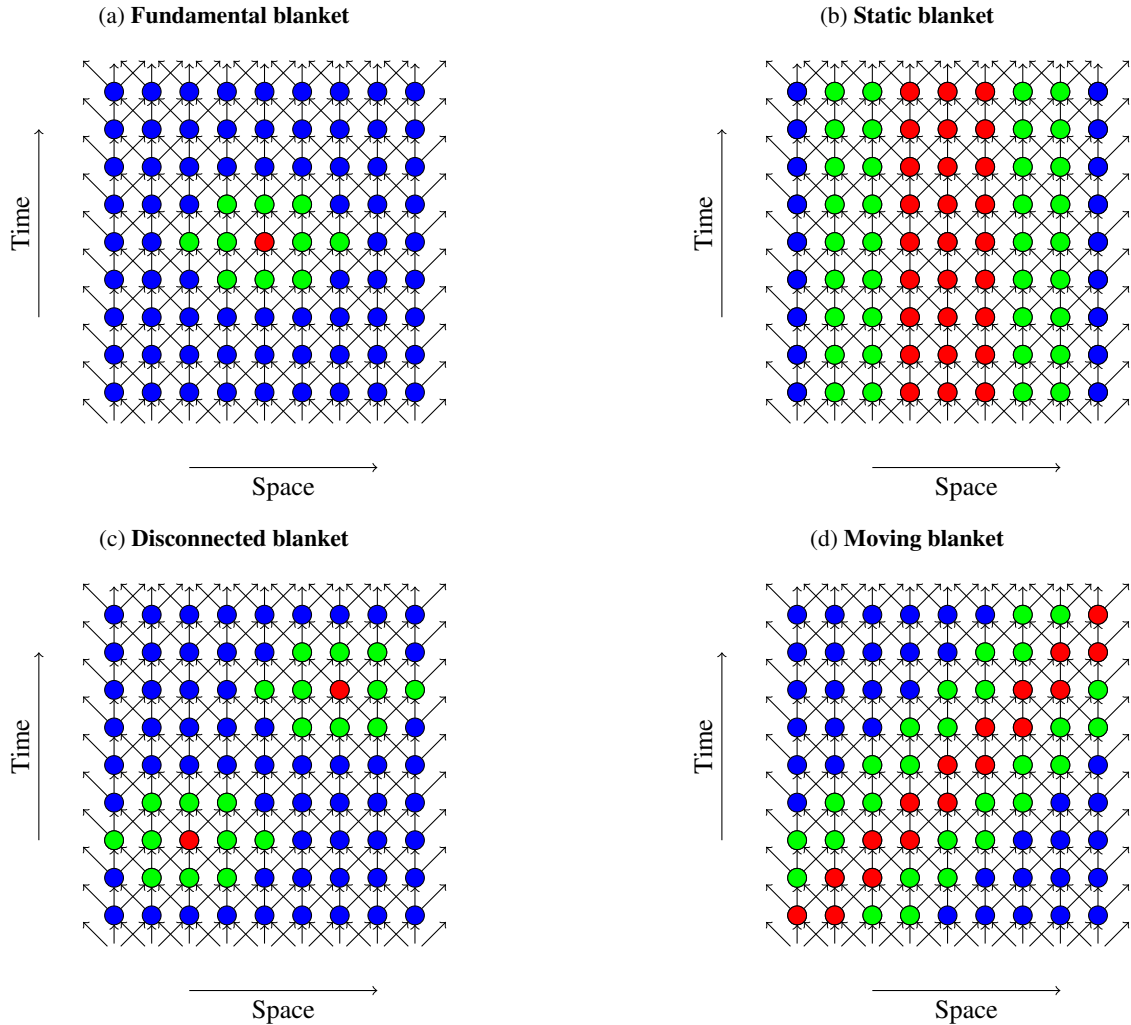
\begin{figure}
  \begin{subfigure}[t]{0.45\textwidth}
    \centering
    \subcaption{\textbf{Fundamental blanket}}
    \vspace{1em}
    \begin{tikzpicture}[scale=0.5]
      % Nodes
      \foreach \y in {0,...,8} {
        \foreach \x in {0,...,8} {
            \draw[draw=black,fill=blue] (\x,\y) circle (0.25);
        }
      }
      \draw[draw=black,fill=red] (4,4) circle (0.25);
      \foreach \x in {3,...,5} {
            \draw[draw=black,fill=green] (\x,3) circle (0.25);
            \draw[draw=black,fill=green] (\x,5) circle (0.25);
        }    
      \foreach \x in {2,...,3}{
        \draw[draw=black,fill=green] (\x,4) circle (0.25);
        }
      \foreach \x in {5,...,6}{
        \draw[draw=black,fill=green] (\x,4) circle (0.25);
      }
      
      % Connections
      \foreach \x in {0,...,8} {
        \foreach \y in {-1,...,8} {
          \draw[->,thin] (\x,{\y+0.25}) -- (\x,{\y+0.75});
          \draw[->,thin] ({\x+0.7071*0.25},{\y+0.25*0.7071}) -- (\x+1-0.7071*0.25,{\y+1-0.25*0.7071});
          \draw[->,thin] ({\x-0.7071*0.25},{\y+0.25*0.7071}) -- (\x-1+0.7071*0.25,{\y+1-0.25*0.7071});
        }
      }
    
        % X-axis arrow
      \draw[->,thin] (2,-2) -- (7,-2) node[midway,below] {Space};
      
      % Time axis arrow
      \draw[->,thin] (-2,2) -- (-2,7) node[midway,above,rotate=90] {Time};
    \end{tikzpicture}
  \end{subfigure}
  \hfill
  \begin{subfigure}[t]{0.45\textwidth}
    \centering
    \subcaption{\textbf{Static blanket}}
    \vspace{1em}
    \begin{tikzpicture}[scale=0.5]
      % Nodes
      \foreach \y in {0,...,8} {
        \foreach \x in {0,...,8} {
            \draw[draw=black,fill=blue] (\x,\y) circle (0.25);
        }
      }
      \draw[draw=black,fill=red] (4,4) circle (0.25);
      \foreach \y in {0,...,8}{
        \foreach \x in {1,...,2} {
            \draw[draw=black,fill=green] (\x,\y) circle (0.25);
        }
        \foreach \x in {3,...,5} {
            \draw[draw=black,fill=red] (\x,\y) circle (0.25);
        }    
        \foreach \x in {6,...,7} {
            \draw[draw=black,fill=green] (\x,\y) circle (0.25);
        }        
        \draw[draw=black,fill=blue] (0,\y) circle (0.25);
        \draw[draw=black,fill=blue] (8,\y) circle (0.25);
      }
    
      % Connections
      \foreach \x in {0,...,8} {
        \foreach \y in {-1,...,8} {
          \draw[->,thin] (\x,{\y+0.25}) -- (\x,{\y+0.75});
          \draw[->,thin] ({\x+0.7071*0.25},{\y+0.25*0.7071}) -- (\x+1-0.7071*0.25,{\y+1-0.25*0.7071});
          \draw[->,thin] ({\x-0.7071*0.25},{\y+0.25*0.7071}) -- (\x-1+0.7071*0.25,{\y+1-0.25*0.7071});
        }
      }
    
        % X-axis arrow
      \draw[->,thin] (2,-2) -- (7,-2) node[midway,below] {Space};
      
      % Time axis arrow
      \draw[->,thin] (-2,2) -- (-2,7) node[midway,above,rotate=90] {Time};
    \end{tikzpicture}
  \end{subfigure}
  \vspace{1em}
  \begin{subfigure}[t]{0.45\textwidth}
    \centering
    \subcaption{\textbf{Disconnected blanket}}
    \vspace{1em}
    \begin{tikzpicture}[scale=0.5]
      % Nodes
      \foreach \y in {0,...,8} {
        \foreach \x in {0,...,8} {
            \draw[draw=black,fill=blue] (\x,\y) circle (0.25);
        }
      }
      \draw[draw=black,fill=red] (4-2,4-2) circle (0.25);
      \foreach \x in {3,...,5} {
            \draw[draw=black,fill=green] (\x-2,3-2) circle (0.25);
            \draw[draw=black,fill=green] (\x-2,5-2) circle (0.25);
        }    
      \foreach \x in {2,...,3}{
        \draw[draw=black,fill=green] (\x-2,4-2) circle (0.25);
        }
      \foreach \x in {5,...,6}{
        \draw[draw=black,fill=green] (\x-2,4-2) circle (0.25);
      }

      \draw[draw=black,fill=red] (4+2,4+2) circle (0.25);
      \foreach \x in {3,...,5} {
            \draw[draw=black,fill=green] (\x+2,3+2) circle (0.25);
            \draw[draw=black,fill=green] (\x+2,5+2) circle (0.25);
        }    
      \foreach \x in {2,...,3}{
        \draw[draw=black,fill=green] (\x+2,4+2) circle (0.25);
        }
      \foreach \x in {5,...,6}{
        \draw[draw=black,fill=green] (\x+2,4+2) circle (0.25);
      }

      % Connections
      \foreach \x in {0,...,8} {
        \foreach \y in {-1,...,8} {
          \draw[->,thin] (\x,{\y+0.25}) -- (\x,{\y+0.75});
          \draw[->,thin] ({\x+0.7071*0.25},{\y+0.25*0.7071}) -- (\x+1-0.7071*0.25,{\y+1-0.25*0.7071});
          \draw[->,thin] ({\x-0.7071*0.25},{\y+0.25*0.7071}) -- (\x-1+0.7071*0.25,{\y+1-0.25*0.7071});
        }
      }
    
        % X-axis arrow
      \draw[->,thin] (2,-2) -- (7,-2) node[midway,below] {Space};
      
      % Time axis arrow
      \draw[->,thin] (-2,2) -- (-2,7) node[midway,above,rotate=90] {Time};
    \end{tikzpicture}
    % \begin{tikzpicture}[scale=0.5]
    %   % Nodes
    %   \foreach \y in {0,...,8} {
    %     \foreach \x in {0,...,8} {
    %         \draw[draw=black,fill=blue] (\x,\y) circle (0.25);
    %     }
    %   }
    %   \foreach \y in {2,...,8} {
    %     \draw[draw=black,fill=red] (4,\y) circle (0.25);
    %   }
    %   \foreach \x in {3,...,5}{
    %         \draw[draw=black,fill=green] ({\x},1) circle (0.25);
    %     }
    %   \foreach \x in {2,...,3}{
    %         \draw[draw=black,fill=green] ({\x},2) circle (0.25);
    %     }
    %   \foreach \x in {5,...,6}{
    %         \draw[draw=black,fill=green] ({\x},2) circle (0.25);
    %     }
    %   \foreach \x in {1,...,3}{
    %         \draw[draw=black,fill=green] ({\x},3) circle (0.25);
    %     }
    %   \foreach \x in {5,...,7}{
    %         \draw[draw=black,fill=green] ({\x},3) circle (0.25);
    %     }
    %   \foreach \y in {4,...,8}{
    %     \foreach \x in {0,...,3}{
    %         \draw[draw=black,fill=green] (\x,\y) circle (0.25);
    %     }
    %     \foreach \x in {5,...,8}{
    %         \draw[draw=black,fill=green] (\x,\y) circle (0.25);
    %     }
    %   }
    %   % Connections
    %   \foreach \x in {0,...,8} {
    %     \foreach \y in {-1,...,8} {
    %       \draw[->,thin] (\x,{\y+0.25}) -- (\x,{\y+0.75});
    %       \draw[->,thin] ({\x+0.7071*0.25},{\y+0.25*0.7071}) -- (\x+1-0.7071*0.25,{\y+1-0.25*0.7071});
    %       \draw[->,thin] ({\x-0.7071*0.25},{\y+0.25*0.7071}) -- (\x-1+0.7071*0.25,{\y+1-0.25*0.7071});
    %     }
    %   }
    
    %     % X-axis arrow
    %   \draw[->,thin] (2,-2) -- (7,-2) node[midway,below] {Space};
      
    %   % Time axis arrow
    %   \draw[->,thin] (-2,2) -- (-2,7) node[midway,above,rotate=90] {Time};
    % \end{tikzpicture}
  \end{subfigure}
  \hfill
  \begin{subfigure}[t]{0.45\textwidth}
    \centering
    \subcaption{\textbf{Moving blanket}}
    \vspace{1em}
    \begin{tikzpicture}[scale=0.5]
      % Nodes
      \foreach \y in {0,...,8} {
        \foreach \x in {0,...,8} {
            \draw[draw=black,fill=blue] (\x,\y) circle (0.25);
        }
      }
      \foreach \y in {0,...,8}{
        \draw[draw=black,fill=red] (\y,\y) circle (0.25);
      }
    
      \foreach \y in {0,...,3}{
            \draw[draw=black,fill=red] ({2*\y+1},{2*\y}) circle (0.25);
            \draw[draw=black,fill=red] ({2*\y+2},{2*\y+1}) circle (0.25);
      }    
      \foreach \y in {0,...,7}{
            \draw[draw=black,fill=green] (\y,{\y+1}) circle (0.25);        
      }
      \foreach \y in {0,...,6}{
            \draw[draw=black,fill=green] (\y,{\y+2}) circle (0.25);        
      }
      \foreach \x in {2,...,8}{
            \draw[draw=black,fill=green] (\x,{\x-2}) circle (0.25);        
      }
      \foreach \x in {3,...,8}{
            \draw[draw=black,fill=green] (\x,{\x-3}) circle (0.25);        
      }
    
      % Connections
      \foreach \x in {0,...,8} {
        \foreach \y in {-1,...,8} {
          \draw[->,thin] (\x,{\y+0.25}) -- (\x,{\y+0.75});
          \draw[->,thin] ({\x+0.7071*0.25},{\y+0.25*0.7071}) -- (\x+1-0.7071*0.25,{\y+1-0.25*0.7071});
          \draw[->,thin] ({\x-0.7071*0.25},{\y+0.25*0.7071}) -- (\x-1+0.7071*0.25,{\y+1-0.25*0.7071});
        }
      }
    
        % X-axis arrow
      \draw[->,thin] (2,-2) -- (7,-2) node[midway,below] {Space};
      
      % Time axis arrow
      \draw[->,thin] (-2,2) -- (-2,7) node[midway,above,rotate=90] {Time};
    \end{tikzpicture}
  \end{subfigure}

  \caption{In a spatially connected domain with nearest neighbor connectivity, the blanket of a single node at time $t$ (a) consists of its parents (the three adjacent nodes at the previous time step), its children (the three adjacent nodes at the next time step), and its co-parents (which consist of the two nodes immediately to the left and right). This implies that gradients can always be computed on the boundary. A static or stationary blanket (b) is defined as having a spatial location that does not change. The intersection of any two blankets makes up blanket allowing for blankets to be disconnected (c). Traveling waves are associated with moving blankets (d). All of these blankets (and many, many more) are present solely by virtue of the topology of the network independent of its dynamics.}
  \label{fig:MB}
\end{figure}

Having established that sensible boundaries correspond to Markov blankets, it is tempting to conclude that any collection of elements that has a Markov blanket corresponds to an object. Unfortunately, the definition of a Markov blanket is too expansive to be of practical use, since every snapshot of every microscopic element has a blanket (Fig \ref{fig:MB}a); see \cite{bruineberg2022emperors} for critical discussion. For this reason, much of the FEP literature has adopted the additional principle for ``thing-ness'': namely, stationarity \cite{dacosta2021bayesian}. That is, the blanket and object must be a proper subset of the system as a whole, and the elements that make up the blanket and object must not change with time \cite{fristonparticular}. This assumption eliminates an arbitrary fluid element from consideration, since matter can freely pass into and out of the element. As it does, the matter transitions from being inside to outside the object by passing through the blanket, and thus, the blanket is not stationary with respect to the elements that make up the fluid. But this rules out the possibility of modeling all sorts of interesting systems, indeed perhaps the most interesting ones, which have dynamic boundaries and which experience material turnover. 

Here, we argue that this stationary formulation of boundaries is both too restrictive and not restrictive enough for our purposes. On the one hand, it is too restrictive, because blanket stationarity prohibits one from concluding that a flame, or any traveling wave, for that matter, is an object. This is because the elements of the medium that make up a traveling wave change as the wave moves through matter. On the other hand, it is not restrictive enough, because it implies that any arbitrary connected subset of elements correspond to an object. In the flame example above, this means that any segment of the original fuse, down which the flame burns, must count as a thing, independent of the temperature profile, reaction rate, and their dynamics.

What this example makes clear is that a more general definition of an object or thing should be able to accommodate the notion of a blanket that can move or change, and should incorporate some aspects of the dynamics of the system as a whole. Moreover, the inclusion of dynamic blankets results in a massive expansion of the number of blankets present in any given system. It is clear that what is needed is an \textit{additional principle} that allows us to select those blankets that correspond to the things that we would like to call things. Since the goal is to sensibly carve the world at its joints, an obvious choice for such a principle is Occam's razor, applied to the global dynamics. We implement this new principle by seeking low-dimensional dynamical systems with dynamic Markov blanket structure and employ a Bayesian modeling approach to instantiate Occam's razor.

\section{A dynamic Markov blanket detection algorithm}

We now present, in general form, a probabilistic generative model with dynamic Markov blanket structure that can be inverted to identify Markov blankets and well as classify objects into types according to their blanket statistics and dynamics. Markov blanket detection is, in general, a difficult problem. Even in static settings, it is NP-hard \cite{glymour2019review, wang2020towards}. This is because, even in a situation where the blankets are stationary, the number of Markov blankets in a given system grows combinatorially with the number of system components. Allowing for blankets to be dynamic only makes things worse. 

We sidestep this issue by taking inspiration from macroscopic physical discovery, which focuses on discovering low-dimensional dynamics that summarize high-dimensional systems. Specifically, we propose a class of dimensionality reduction algorithms that partition high-dimensional dynamical systems into subsystems that have Markov blanket structure. This is accomplished via the presumption that low-dimensional latent dynamics have Markov blanket structure and that each element in the original high-dimensional observation space is driven by just one of the low-dimensional latents. For example, in the case where a single object is assumed to be present, we seek a set of state variables that represents the environment ($s$) and two sets of state variables for each object: one that summarizes the collective behavior of the elements that belong to the boundary ($b$) and another that describes the collective behavior of the elements that are inside the boundary ($z$). We leave open the possibility that the set of internal elements may be empty. 

\begin{figure}
    \centering
    \begin{subfigure}{0.4\textwidth}
        \centering
        \includegraphics[width=\linewidth]{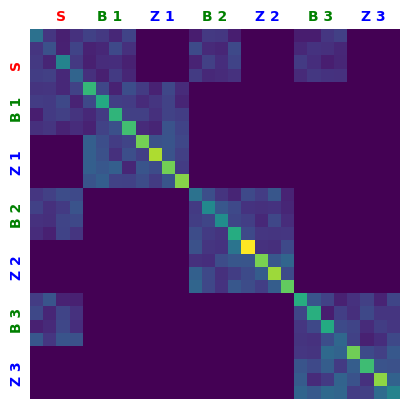} % Path to the first image
        \caption{Transition matrix ($A$ or $T$)}
    \end{subfigure}
    \hfill
    \begin{subfigure}{0.4\textwidth}
        \centering
        \includegraphics[width=\linewidth]{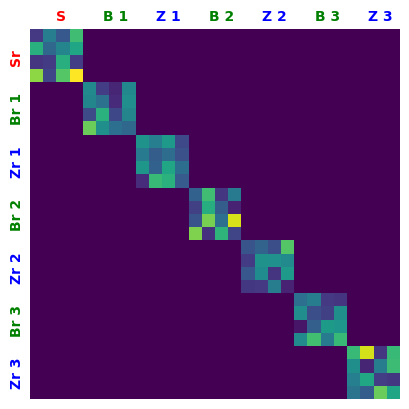} % Path to the second image
        \caption{Observation tensor $B[:,\mathrm{roles},\mathrm{latents}]$}
    \end{subfigure}
    \caption{(a) Constraints on transition matrices and observation tensors for $N=3$ objects. (b) The observation model, under the assumption that there are 4 roles per latent and a 4 dimensions in each observation vector $y_i(t)$.}
    \label{fig:blanketmasks}
\end{figure}

We design enough flexibility into the model to discover objects that have non-stationary blankets by directly modeling the boundary, $\Sigma_B(t)$, as a latent assignment variable. This is accomplished by instantiating a dynamic attention mechanism that probabilistically assigns a label to each microscopic element or measurement. These labels indicate whether that element is an internal, external, or blanket element. Transitions between these labels obey the usual rules for Markov blankets, i.e., label transitions from internal to external are prohibited and label transition probabilities depend only on the macroscopic blanket variables. When multiple objects are present, we allow for the possibility that there are a set of macroscopic variable for the boundary of each object ($b_n$) as well as its internal state ($z_n$), $n=1\cdots N$ and a single environment variable $s$. Observed data, denoted $y_i(t)\in \mathbb{R}^D$, are assumed to be a relatively fine-grained measurement of the activity. Associated with each observation index $i$ is one of the discrete time dependent labels $\omega_i(t) \in \left\{S,B_n,Z_n\right\}$. These labels identify the boundary of each object and determine which of the macroscopic variables influences the associated microscopic observation. Specifically, the label associated with each measurement determines the conditional independence relationship:

\begin{equation}
    p \left(y_i(t) | \omega_i(t) = Z_n, s(t), \{b(_nt),z_n(t)\} \right) = p\left( y_i(t) | \omega_i(t) = Z_n, z_n(t)\right)
\end{equation}

\noindent with similar equations holding for the cases where $\omega_i(t)$ is $S$ or $B_n$. Though written in generative form, this observation model corresponds to a noisy non-linear projection from observations to target macroscopic variables, modulated by the assignment variables. The dynamics of both the macroscopic variables and the transition probabilities of the assignment variables are constrained to obey Markov blanket structure, with the added restriction that the transition probabilities for the labels depend only on the boundary variables. 

\begin{equation}
\frac{d\mathbf{p}_i}{dt} = \mathbf{T}(\{b_n\})\mathbf{p}_i
\end{equation}

where $\mathbf{T}(\cdot)$ is a 1+2n x 1+2n matrix, constrained such that $\mathbf{T}_{SZ_n}(\{b_n\} = \mathbf{T}_{Z_nS}(\{b_n\}) = 0$ prohibiting element labels from transitioning directly from object to environment. Similarly, the dynamics of the latent variables are constrained such that the Jacobian of the global dynamics only allows $z_n$-$b_n$ and $b_n$-$s$ interactions. See Fig. \ref{fig:blanketmasks}a. 

\begin{figure}
\centering
\begin{tikzpicture}[node distance=2.5cm]
  % Define nodes
  \node[draw, circle, minimum size=1cm] (s) {$s^t$};
  \node[draw, circle, below of=s, minimum size=1cm] (b) {$b^t$};
  \node[draw, circle, below of=b, minimum size=1cm] (z) {$z^t$};
  \node[draw, circle, below of=z, fill=lightgray, minimum size=1cm] (y) {$y_i^t$};
  \node[draw, circle, below of=y, minimum size=1cm] (lambda) {$\omega_i^t$};

  % Add edges

  % Add replicated nodes
  \node[draw, circle, minimum size=1cm, left of=s] (s-1) {$s^{t-1}$};
  \node[draw, circle, below of=s-1, minimum size=1cm] (b-1) {$b^{t-1}$};
  \node[draw, circle, below of=b-1, minimum size=1cm] (z-1) {$z^{t-1}$};
  \node[draw, circle, below of=z-1, fill=lightgray, minimum size=1cm] (y-1) {$y_i^{t-1}$};
  \node[draw, circle, below of=y-1, minimum size=1cm] (lambda-1) {$\omega_i^{t-1}$};

  \node[draw, circle, minimum size=1cm, right of=s] (s+1) {$s^{t+1}$};
  \node[draw, circle, below of=s+1, minimum size=1cm] (b+1) {$b^{t+1}$};
  \node[draw, circle, below of=b+1, minimum size=1cm] (z+1) {$z^{t+1}$};
  \node[draw, circle, below of=z+1, fill=lightgray, minimum size=1cm] (y+1) {$y_i^{t+1}$};
  \node[draw, circle, below of=y+1, minimum size=1cm] (lambda+1) {$\omega_i^{t+1}$};

  % Add arrows
  \draw[->] (s) to[out=-45, in=45, distance=1cm] (y);
  \draw[->] (b) to[out=-45, in=45, distance=1cm] (y);
  \draw[->] (z) -- (y);
  \draw[->] (s+1) to[out=-45, in=45, distance=1cm] (y+1);
  \draw[->] (b+1) to[out=-45, in=45, distance=1cm] (y+1);
  \draw[->] (z+1) -- (y+1);
  \draw[->] (s-1) to[out=-45, in=45, distance=1cm] (y-1);
  \draw[->] (b-1) to[out=-45, in=45, distance=1cm] (y-1);
  \draw[->] (z-1) -- (y-1);
  \draw[->] (lambda-1) -- (lambda);
  \draw[->] (lambda) -- (lambda+1);
  \draw[->] (lambda) -- (y);
  \draw[->] (lambda-1) -- (y-1);
  \draw[->] (lambda+1) -- (y+1);

  \draw[->,dashed] (b-1) to[out=-45, in=135, distance=1cm] (lambda);
  \draw[->,dashed] (b) to[out=-45, in=135, distance=1cm] (lambda+1);
  \draw[->,dashed] (b+1) to[out=-45, in=135, distance=1cm] ++(1.5,-7);
  \draw[->,dashed] ($(lambda-1) + (-1.5, 7)$) to[out=-45, in=135, distance=1cm] (lambda-1);

  \draw[->] (s) to[out=-45, in=135, distance=1cm] (b+1);
  \draw[->] (s) -- (s+1);
  \draw[->] (b) to[out=45, in=-135, distance=1cm] (s+1);
  \draw[->] (b) to[out=-45, in=135, distance=1cm] (z+1);
  \draw[->] (b) -- (b+1);
  \draw[->] (z) to[out=45, in=-135, distance=1cm] (b+1);
  \draw[->] (z) -- (z+1);

  \draw[->] (s-1) -- (s);
  \draw[->] (s-1) to[out=-45, in=135, distance=1cm] (b);
  \draw[->] (z-1) -- (z);
  \draw[->] (z-1) to[out=45, in=-135, distance=1cm] (b);
  \draw[->] (b-1) -- (b);
  \draw[->] (b-1) to[out=45, in=-135, distance=1cm] (s);
  \draw[->] (b-1) to[out=-45, in=135, distance=1cm] (z);

  \draw[->] (lambda-1) -- (lambda);
  \draw[->] (lambda-1) -- (y-1);
  \draw[->] (lambda) -- (lambda+1);
  \draw[->] (lambda+1) -- (y+1);

  \draw[->] (s+1) -- ++(1.5,0);
  \draw[->] (s-1) ++(-1.5,0) -- (s-1);
  \draw[->] (b+1) -- ++(1.5,0);
  \draw[->] (b-1) ++(-1.5,0) -- (b-1);
  \draw[->] (z+1) -- ++(1.5,0);
  \draw[->] (z-1) ++(-1.5,0) -- (z-1);
  \draw[->] (lambda+1) -- ++(1.5,0);
  \draw[->] (lambda-1) ++(-1.5,0) -- (lambda-1);
  
  \node[draw, rectangle, minimum height=4.5cm, minimum width=7cm, yshift=-1.5cm, label={[anchor=south]south:{$i=1,2,\ldots,N$}}, fit=(y)] (plate) {};

\end{tikzpicture}
\caption{Generative model for factorial hidden Markov model with Markov blanket structure. The dashed lines determine how the macroscopic blanket variables affects the evolution of the blanket labels. This breaks the factorial structure, increasing the complexity of inference. However, when simulating data for algorithm testing purposes, these connections help maintain the stability or permanence of objects.}
\label{fig:dmbdGMfig}
\end{figure}
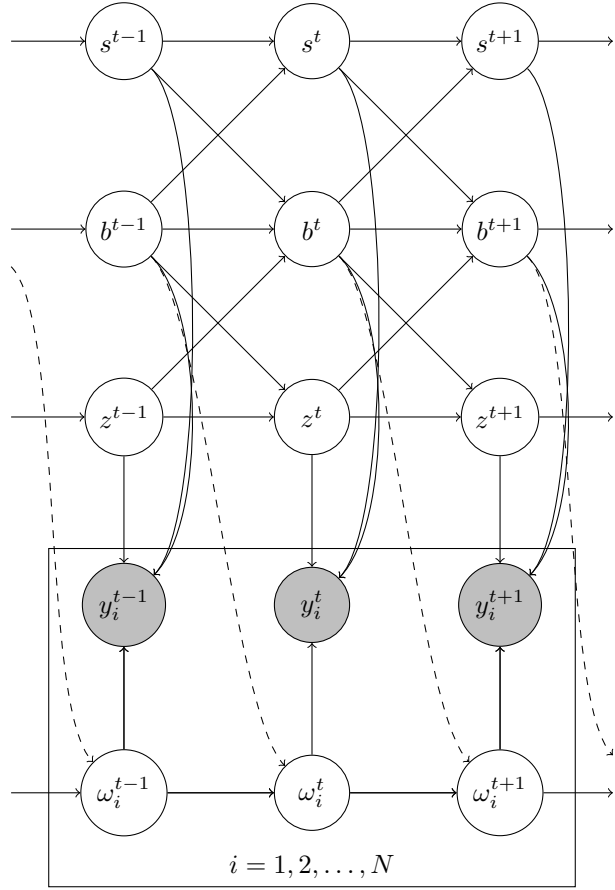

This constitutes the general form of this kind class of blanket discovery algorithms. In order to convert this general dynamic Markov blanket model into a more tractable form, a few additional simplifying assumptions are required. For example, non-linear dynamics could be implemented via recurrent switching dynamics system and discovered through variational inference\cite{linderman17rslds} or specified by a neural network and learned via gradient descent\cite{raissi2019pinn}. Here, we will assume simple linear dynamics and instantiate the non-linear observational model using a switching linear transform. We will also assume that transition probabilities for the assignment variables are \textit{a priori} independent of the macroscopic latents. This leads to a factorial Hidden Markov Model (HMM) that mixes discrete and continuous latent variables, with the unique feature that the labels associated with every observational node $i$ has its own discrete HMM. See Fig. \ref{fig:dmbdGMfig}.

For a single object this linear model is given by

\begin{eqnarray}
s',b',z'|s,b,z  & \sim  & Normal\left(A[s,b,z] + B,\Sigma_{sbz}\right) \\
\omega_i' & \sim & Categorical\left( T\omega_i \right)  \\
y_i & \sim & Normal\left(C_{\omega_i}[s,b,z] + D_{\omega_i},\Sigma_{\omega_i}\right)
\end{eqnarray}

where the matrix $A$ is constrained to have Markov blanket structure by placing blocks of zeros in the upper right and lower left corners. Similarly, the means of $C_S$, $C_B$, and $C_Z$ are constrained to have blocks of zeros so that observations only depend upon one of the three continuous latents. These constraints are imposed on the mean via the action of Lagrange multipliers. We reduce the degeneracy of the model by assuming that $\Sigma_{sbz}$ is diagonal, with unit trace imposed on the posterior mean, also via the action of Lagrange multipliers. This further encourages the discovery of low-dimensional macroscopic dynamics via the action of the automatic Occam`s razor effect of Bayesian inference.  

In order to model the non-linear observation model, we expand the domain of the discrete latent assignment variables to include ``roles'' associated with each of the labels, $S,B,Z$. This effectively instantiates a hierarchical hidden Markov model (HHMM) for the label assignment variables, where the blanket structure is enforced at the highest level of the hierarchy, and the lowest level represents a mixture of linear experts model to describe the relationship between the targeted macroscopic variable and the observation. This is depicted in Fig. \ref{fig:blanketmasks}b. Note that we could have used the same trick for the dynamics, effectively implementing a recurrent switching dynamical system, but rationalized linear dynamics with a non-linear observation model, as an instantiation of the Koopman embedding trick \cite{koopman1931hamiltonian}

\subsection{Inference and learning}

We approximate Bayesian inference using variational Bayesian expectation maximization (VBEM), with a posterior that factorizes over parameters and the two classes of latents, i.e.

\begin{equation}
q(s,b,z,\omega,\Theta) = q_{sbz}(s,b,z)q_{\omega}(\omega)q_\Theta(\Theta)
\end{equation}

where $\Theta = \{A,B,C_\omega,D_\omega, \Sigma_{sbz}^{-1}, \Sigma_{S,B,Z}^{-1}, T\}$. This leads to an inference algorithm that iteratively updates $q_{sbz}(s,b,z)$ and $q_{\omega}(\omega)$ until convergence, i.e.,

\begin{eqnarray}
\log q_{\omega}(\omega) & = & \left< \log p(s,b,z,\omega|\Theta)\right>_{p(s,b,z)p(\Theta)} \\
\log q_{sbz}(s,b,z) & = & \left< \log p(s,b,z,\omega|\Theta) \right>_{p(\omega)p(\Theta)}
\end{eqnarray}

This follows the attend, infer, repeat (AIR) paradigm \cite{eslami2016attend}, where the posterior updates over the assignment variable trajectories, $\omega_i(t)$, corresponds to the attentional update, which is then followed by jointly inferring $s$, $b$, and $z$, and then repeating prior to updating the posterior distributions over the parameters. Note that unrolling these two steps into a neural network implementation leads to a variation on the powerful transformer architecture \cite{vaswani2017attention}, but with added structure that ties attention variables across layers of the network implementation.  

Note that we do not factorize $q_{sbz}(\cdot)$ or $q_{\omega}(\cdot)$ over time, but instead implement a forward-backward scheme that performs exact computation of the posterior distribution over the latent variables\cite{beal2003variational}. Conditional conjugacy allows us to use a matrix normal gamma distribution for $q(A,B,\Sigma^{-1})$, a matrix normal inverse Wishart for $q(C_\omega,D_\omega, \Sigma^{-1}_\omega)$, Dirichlet for rows of $T$, and normal inverse Wishart distributions and Dirichlet distributions for the relevant initial conditions. Coordinate update rules for the natural parameters of the posteriors over $\Theta$ were augmented by a learning rate parameter, typically set to 0.5, in order to enable parameter learning on minibatches of trajectories and reap the benefits of stochastic natural gradient descent \cite{jones2024bong}. 

\section{Results}

We present simple numerical experiments that demonstrate how a dynamic Markov blanket detection algorithm sensibly labels the components of Newton's cradle, a burning fuse, the Lorenz attractor, and an artificial life simulation. Inference and learning for this model are implemented using a custom message passing framework, specifically designed to exploit conditional conjugacy and stochastic coordinate ascent \cite{hoffman2013stochastic}. The code for the dynamic Markov blanket detection algorithm and the VBEM inference modules upon which it is built can be found at \url{https://github.com/bayesianempirimancer/pyDMBD}. All simulations are trained on the complete data set, with a learning rate set to 0.5 for 50 training epochs. This is done a minimum of 10 times. Results shown are from the computational run that achieves the largest expected lower bound (ELBO) on the log likelihood. 

\subsection{Newton's cradle}

The simulation of Newton's cradle consists of 5 balls of equal size and shape that are suspended from strings, which are separated by a distance equal to the diameter of any ball. At rest, the balls hang together, just barely touching. When the leftmost ball is given an initial position away from the other 4, it swings down, halts its motion at its resting position, and transfers its momentum through to the other balls. These move in sequence and similarly transfer their momentum up to the rightmost ball, which this swings up and away; before returning, and repeating the process. If two balls are initially perturbed, then two balls will pop out the other side, and so on.  We simulate a Newton's cradle with either zero, one, two, or three balls initially perturbed by the same randomly selected angle between 0 and $3 \pi/2$. All the balls are then randomly perturbed by a small angle difference, with standard deviation $0.1$ radians. 

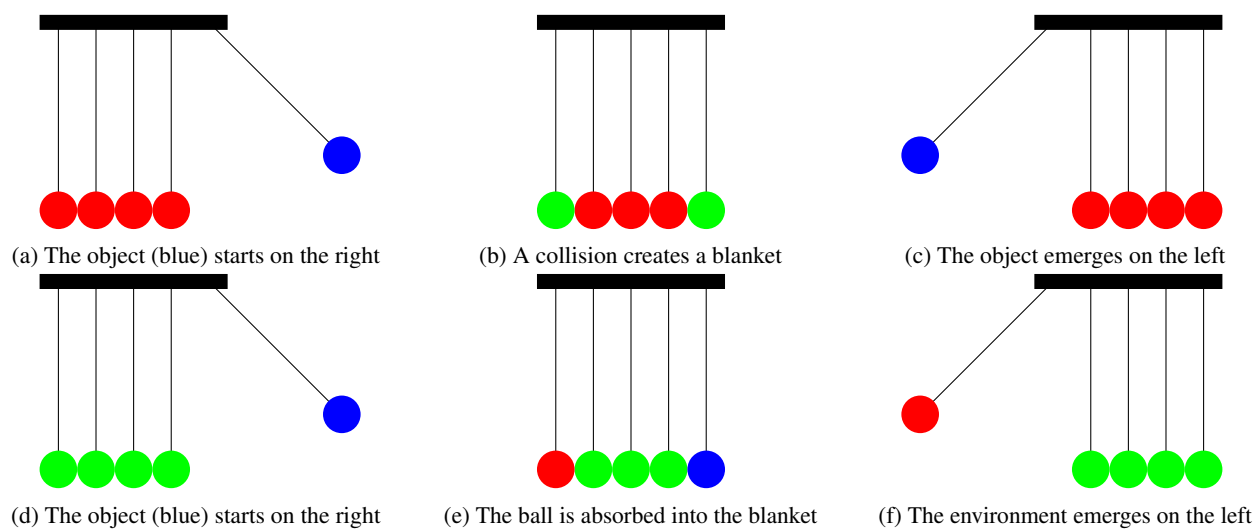
\begin{figure}[ht]
  \centering
  \begin{subfigure}{0.3\textwidth}
    \centering
    \begin{tikzpicture}[scale=0.25]
      % Strings
      \draw[line width=0.2cm] (-5, 0) -- (5, 0);
      \foreach \x in {-4, -2, 0, 2} {
        \draw (\x, 0) -- (\x, -10);
      }
      \draw (4,0) -- (4+10*0.7071,-10*0.7071);
      % Balls
      \foreach \x in {-4, -2, 0, 2} {
        \fill[red] (\x, -10) circle (1);
      }
      \fill[blue] (4+10*0.7071,-10*0.7071) circle(1);  
    \end{tikzpicture}
    \caption{The object (blue) starts on the right}
  \end{subfigure}%
  \hfill
  \begin{subfigure}{0.3\textwidth}
    \centering
    \begin{tikzpicture}[scale=0.25]
      % Strings
      \draw[line width=0.2cm] (-5, 0) -- (5, 0);
      \foreach \x in {-4, -2, 0, 2, 4} {
        \draw (\x, 0) -- (\x, -10);
      }
      \foreach \x in {-2, 0, 2} {
        \fill[red] (\x, -10) circle (1);
      }
      \fill[green] (-4,-10) circle(1);  
      \fill[green] (4,-10) circle(1);  
    \end{tikzpicture}
    \caption{A collision creates a blanket}
  \end{subfigure}%
  \hfill
  \begin{subfigure}{0.3\textwidth}
    \centering
    \begin{tikzpicture}[scale=0.25]
      % Strings
      \draw[line width=0.2cm] (-5, 0) -- (5, 0);
      \foreach \x in {4, -2, 0, 2} {
        \draw (\x, 0) -- (\x, -10);
      }
      \draw (-4,0) -- (-4-10*0.7071,-10*0.7071);
      % Balls
      \foreach \x in {4, -2, 0, 2} {
        \fill[red] (\x, -10) circle (1);
      }
      \fill[blue] (-4-10*0.7071,-10*0.7071) circle(1);  
    \end{tikzpicture}
    \caption{The object emerges on the left}
  \end{subfigure}

  \centering
  \begin{subfigure}{0.3\textwidth}
    \centering
    \begin{tikzpicture}[scale=0.25]
      % Strings
      \draw[line width=0.2cm] (-5, 0) -- (5, 0);
      \foreach \x in {-4, -2, 0, 2} {
        \draw (\x, 0) -- (\x, -10);
      }
      \draw (4,0) -- (4+10*0.7071,-10*0.7071);
      % Balls
      \foreach \x in {-4, -2, 0, 2} {
        \fill[green] (\x, -10) circle (1);
      }
      \fill[blue] (4+10*0.7071,-10*0.7071) circle(1);  
    \end{tikzpicture}
    \caption{The object (blue) starts on the right}
  \end{subfigure}%
  \hfill
  \begin{subfigure}{0.3\textwidth}
    \centering
    \begin{tikzpicture}[scale=0.25]
      % Strings
      \draw[line width=0.2cm] (-5, 0) -- (5, 0);
      \foreach \x in {-4, -2, 0, 2, 4} {
        \draw (\x, 0) -- (\x, -10);
      }
      \foreach \x in {-2, 0, 2} {
        \fill[green] (\x, -10) circle (1);
      }
    \fill[red] (-4,-10) circle(1);  
    \fill[blue] (4,-10) circle(1);  

    \end{tikzpicture}
    \caption{The ball is absorbed into the blanket}
  \end{subfigure}%
  \hfill
  \begin{subfigure}{0.3\textwidth}
    \centering
    \begin{tikzpicture}[scale=0.25]
      % Strings
      \draw[line width=0.2cm] (-5, 0) -- (5, 0);
      \foreach \x in {4, -2, 0, 2} {
        \draw (\x, 0) -- (\x, -10);
      }
      \draw (-4,0) -- (-4-10*0.7071,-10*0.7071);
      % Balls
      \foreach \x in {4, -2, 0, 2} {
        \fill[green] (\x, -10) circle (1);
      }
      \fill[red] (-4-10*0.7071,-10*0.7071) circle(1);  
    \end{tikzpicture}
    \caption{The environment emerges on the left}
  \end{subfigure}
  
  \caption{Newton's Cradle. In (a-c), the ball that makes up the object (blue) collides with the environment (red). Since the force transmitted must pass through the boundary, the dynamic Markov blanket discovery algorithm labels the balls on the periphery as temporarily becoming part of the blanket. The momentum is then passed through to the balls that make up the environment, and is transferred to the ball on the left, which also temporarily becomes part of the blanket, before emerging as the object on the other side. In (d-f), we see another common solution to this problem that is discovered by the dynamic Markov blanket discovery algorithm. In that solution, the blanket consists of the stationary balls. When energy is transferred to the ball on the right, it is part of the object. When energy is transferred to the ball on the left, it is part of the environment.}
  \label{fig:cradle}
\end{figure}

Applied to this data set, a \textit{static} Markov blanket detection algorithm based upon exerted forces \cite{fristonparticular} discovers a single object, centered on the middle ball, regardless of the dynamics of the system, and without regard for the number of initially displaced balls. In contrast, the \textit{dynamic} Markov blanket detection algorithm labels the balls in a manner that is consistent with human intuition: that is, we tend to perceive Newton's cradle dynamics as either as a pendulum or as a set pair of interacting sets of balls, one on the left and one on the right. These two common precepts map precisely onto the two most commonly discovered partition discovered by this simple DMBD algorithm. Specifically, Fig. \ref{fig:cradle} depicts these two solutions, where color indicates label, with green for boundary, and red and blue for object and environment. If two balls or more balls are initially perturbed, then the balls that move together are always given the same assignment. The most commonly discovered solution (Fig. \ref{fig:cradle}(a-c)) labels the object as the moving balls, regardless of which side of the cradle they are on. In this case, the environment label is assigned to the more or less stationary balls. The boundary consists of the balls that very briefly obtain high velocity due to collision. As a result, the boundary is not physically realised most of the time. The second most commonly identified partition locates the boundary as the stationary balls, and labels the object and environment as the moving balls. The ball on one side is always labeled environment and the ball on the other side is labeled object. See Fig \ref{fig:cradle}(d-f). Once again, when the leftmost or rightmost balls are more or less stationary, they become part of the boundary. 

\begin{figure}[ht]
  \centering

  \begin{subfigure}{0.45\textwidth}
    \centering
    \includegraphics[width=\linewidth]{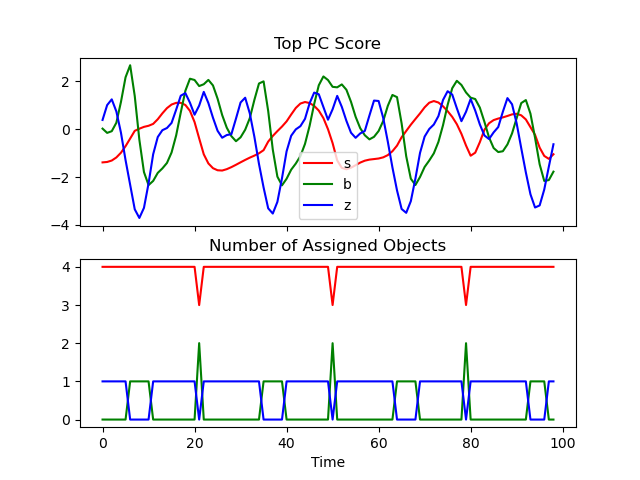}
    \caption{Case 1: Middle balls correspond to the environment and moving balls is the environment. During transition of the object from the left to the right side, there is a brief period when the boundary appears associated with impacts.}
  \end{subfigure}
  \hfill
  \begin{subfigure}{0.45\textwidth}
    \centering
    \includegraphics[width=\linewidth]{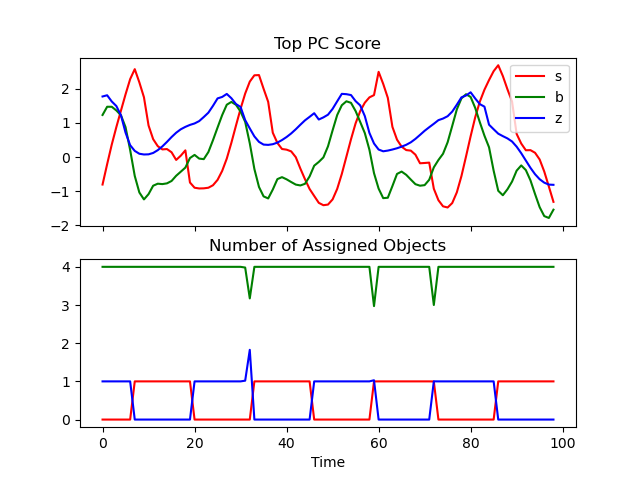}
    \caption{Case 2: The stationary balls are assigned to the boundary.  When the moving ball is on the left, it is labeled as the environment; and when the moving ball is on the right, it is labeled as the object.}
  \end{subfigure}

  \caption{The principal component is computed independently for the most likely path taken for each of the 4 dimensional latents: $s,b,z$. The dynamics of the latent assignments for each ball are summarized by the total number of balls assigned to the environment, blanket, and object. Note that the macroscopic latent variables are highly correlated and persist even when they lack physical realization via assignment to a ball. This is consistent with the notion that internal (and blanket) variables maintain a representation of the environment, consistent with Bayesian mechanics.}
  \label{fig:cradlelatents}
\end{figure}

% \includemovie[
%   poster,
%   toolbar,              % Show player controls
%   label=NewtonsCradle,       % Unique label for referencing
%   text={\small(Click to play)} % Text displayed before playing
% ]{\linewidth}{0.75\linewidth}{videos/cradle.mp4}

\subsection{A burning fuse}

\begin{figure}[ht]
    \centering
    \begin{subfigure}{0.45\textwidth}
    \centering
    \includegraphics[width=\linewidth]{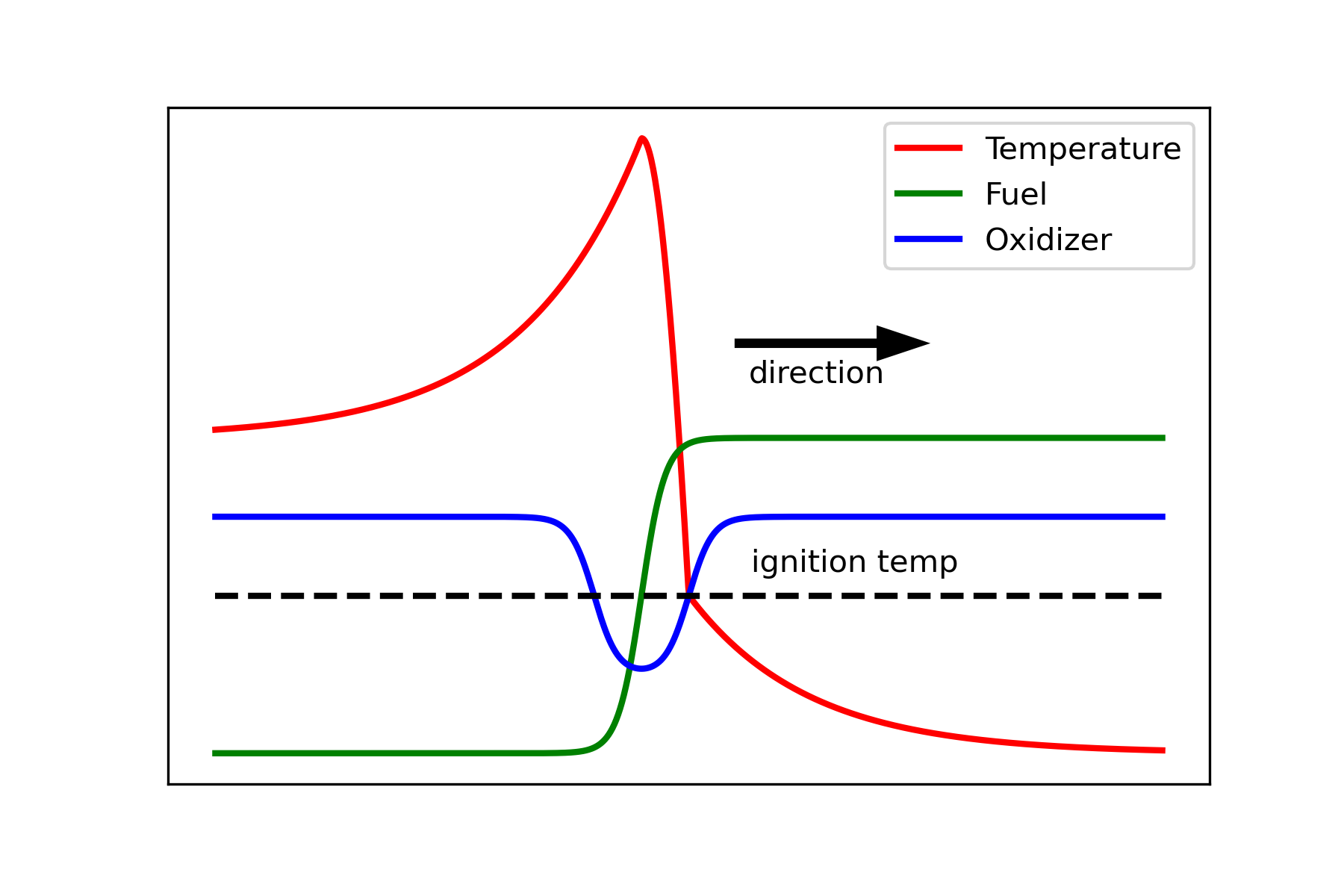}
    \caption{Idealized combustion wave}
    \end{subfigure}
    \hfill
    \begin{subfigure}{0.45\textwidth}
    \centering
    \includegraphics[width=\linewidth]{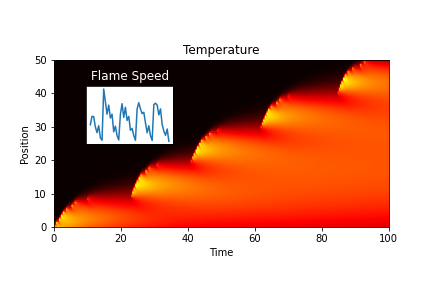} 
    \caption{Fuse in the wind}
    \end{subfigure}
    \caption{}
    \caption{The algorithm is trained on 200 flame trajectories with 4 dimensions per latent and a total of 12 roles. Observables are sampled at 200 even-spaced locations and 800 points in time. Discovered dynamics are roughly 6 dimensional. Typical role usage is about 7 in total. Fig. \ref{fig:flamefit} shows the maximum a posteriori label assignments that the algorithm makes to each node at each point in time, when presented with a flame that goes extinct at approximate $t=120$. Blue represents the flame, red the environment, and green the boundary. The use of roles is critical in this case, because the environment behaves very differently depending upon whether it corresponds to burned versus unburned fuse. Indeed, the blanket only uses two roles: one for each of the front and back of the reaction zone. The object requires only a single role.}
    \label{fig:flame}
\end{figure}

\begin{figure}[ht]
\centering
\begin{subfigure}{0.45\textwidth}
\centering
\includegraphics[width=\linewidth]{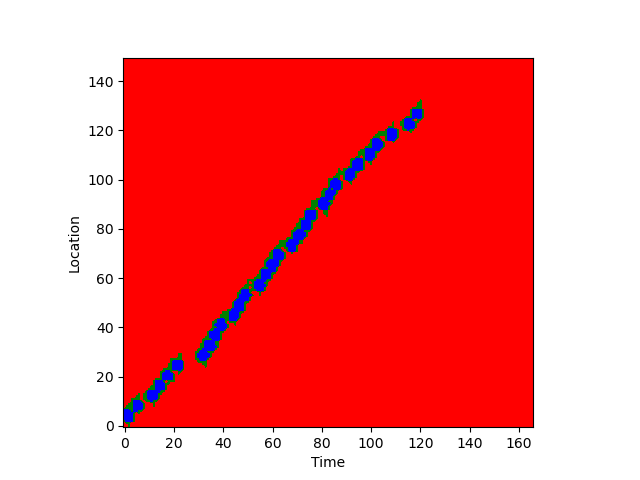}
\caption{Node Assignments: environment in red, blanket in green, object in blue.  Around time t=120 the flame goes extinct and disappears.}
\end{subfigure}
\hfill
\begin{subfigure}{0.45\textwidth}
\centering
\includegraphics[width=\linewidth]{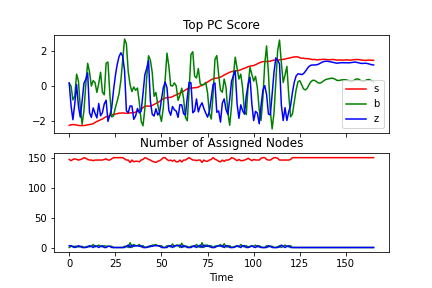} 
\caption{The internal variable (blue) tracks the negative of heat released.  The environment variable tracks flame location. Note that at approximately $t=125$, the flame goes extinct.}
\end{subfigure}
\caption{}
\label{fig:flamefit}
\end{figure}

To demonstrate how this approach can handle flames and traveling waves, we model a combustion front traveling down a one dimensional medium, i.e., a burning fuse. The fuse is modeled as an in-homogeneous medium composed of discrete fuel particles separated by a random distance, with unit mean and variance of 0.02. An oxidizer is modeled as a diffusive process with an inexhaustible source, orthogonal to the fuse. Its effect on combustion is to determine the rate of heat release. Ignition occurs when the particle reaches a critical temperature. The fuse is assumed have constant thermal diffusivity of 1. We select this toy model because it is known to support traveling waves that propagate smoothly, periodically, or chaotically, even in the absence of random perturbations to fuel particle size, location, and oxidizer availability \cite{beck2003nonlinear}. Here, the observable variables, $y_i^t$ consist of the fuel and oxidizer concentrations, as well as the temperature at each location on the fuse, $i$. Broadly speaking, ahead of the combustion front, fuel and oxidizer are plentiful; and behind the combustion front, fuel is absent but oxidizer is plentiful. Inside the combustion front, fuel rapidly drops to zero, oxidizer dips and then returns to a level that corresponds to its current availability, and heat is released (See Fig. \ref{fig:flame}a). We simulate a variety of combustion waves by systematically varying the fuel availability as a function of location and the oxidizer availability as a function of time. The conjunction of these parameters drives variability in rate of heat released as the fuel particles burned. We also manipulate the ignition temperature across a small range of values $(0.4, 0.5)$. This leads to combustion waves that travel at different average speeds, with a range of periodic or random fluctuations. Some combustion waves also go extinct. Fig. \ref{fig:flame}(b) shows an example of a wave that has periodic fluctuations in wave speed. Curiously, the internal variable shows little correlation with object position and seems to only represent heat release \ref{fig:flamefit}. In contrast, it is the environmental variable that is most strongly correlated with flame position. For completeness, we note that while the results shown here are the most common result, the algorithm sometimes converges on less sensible solutions. For example, the object sometimes corresponds to the burned portion behind the combustion wave and the boundary sometimes corresponds to the region in which heat is released. The solutions previously described, however, are associated with the greatest ELBOs and so are strongly preferred by a Bayesian model comparison.

\subsection{Lorenz attractor}

The Lornez attractor also provides a unique test bed for this approach. In this case, there is a single 3 dimensional observation, the $x,y,z$ position of the object. The chaotic dynamics of the system supports two low ($\sim 2$) dimensional attractors and global dynamics that switch between the attracting manifolds. Here, the algorithm discovers what could be called a phase transition, by labeling the observation as part of the ``object'' when it is near to the attractor on the left and part of the ''environment'' when it is near to attractor on the right. The boundary label is sensibly associated with the part of the trajectory that describes the transition between the two attractors. Because of the symmetry associated with object and environment, we could interpret this as modeling a phase transition in which an single object changes identity after passing through a phase transition boundary. 

\begin{figure}[ht]
\centering
\begin{subfigure}{0.33\textwidth}
\centering
\includegraphics[width=\linewidth]{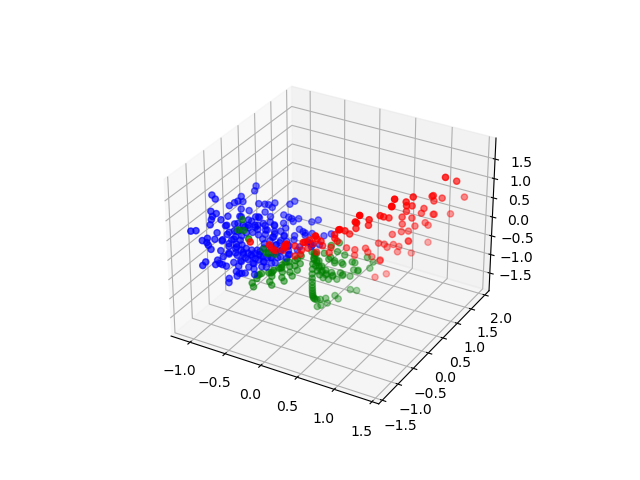}
\caption{Lorenz Attractor.}
\end{subfigure}
\hfill
\begin{subfigure}{0.30\textwidth}
\centering
\includegraphics[width=\linewidth]{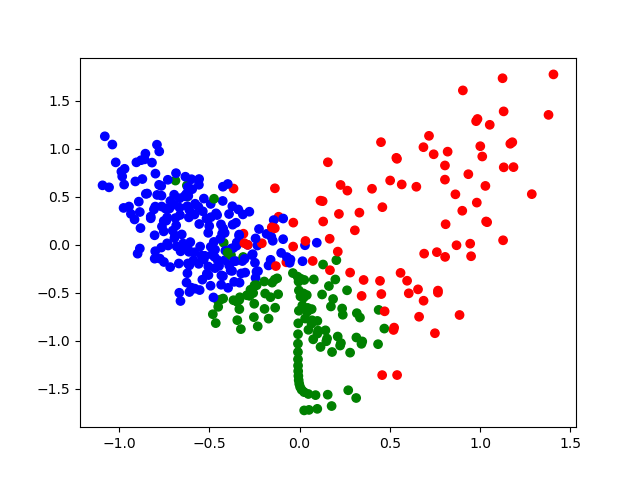}
\caption{Two dimensional projection}
\end{subfigure}
\hfill
\begin{subfigure}{0.30\textwidth}
\centering
\includegraphics[width=\linewidth]{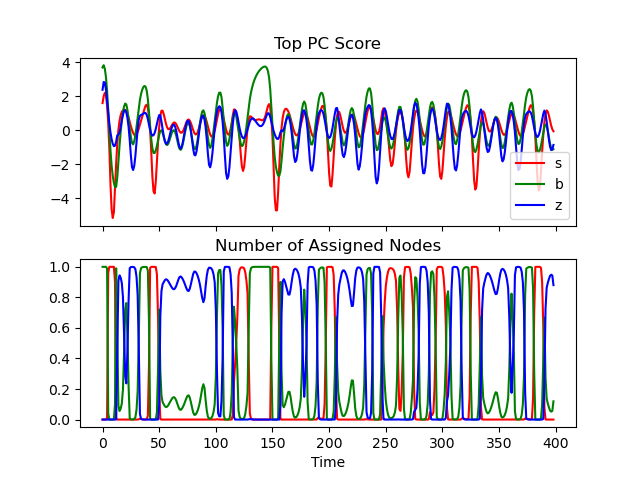}
\caption{Principal components and assignments}
\end{subfigure}

\label{fig:lorenz}
\caption{Red represents the environment, green represents the blanket, blue represents the object. When in the $\sim 2$ dimensional oscillating state, the object is present. Note that, at about $t = 50$, the attractor on the left nearly completely captures the dynamics for an extended period of time.}
\end{figure}

\subsection{Synthetic biology simulation}

\begin{figure}[ht]
  \centering
  \begin{subfigure}{0.3\textwidth}
    \centering
    \includegraphics[width=\linewidth]{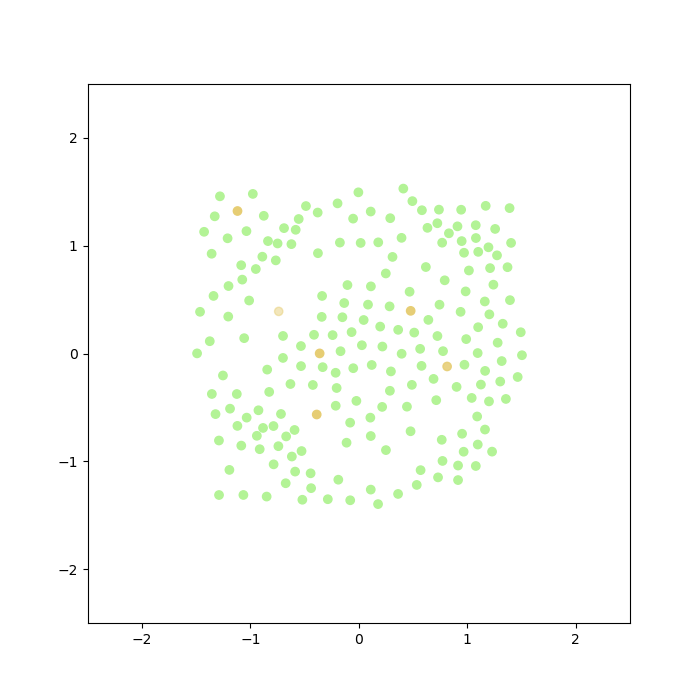}
    \caption{Initial disordered state.}
  \end{subfigure}%
  \hfill
  \begin{subfigure}{0.3\textwidth}
    \centering
    \includegraphics[width=\linewidth]{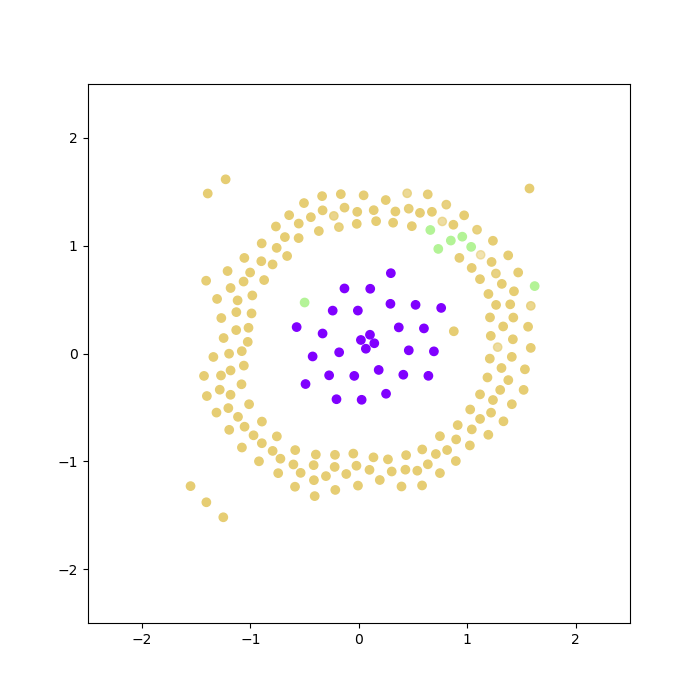}
    \caption{'Cell membrane' forms.}
  \end{subfigure}%
  \hfill
  \begin{subfigure}{0.3\textwidth}
    \centering
    \includegraphics[width=\linewidth]{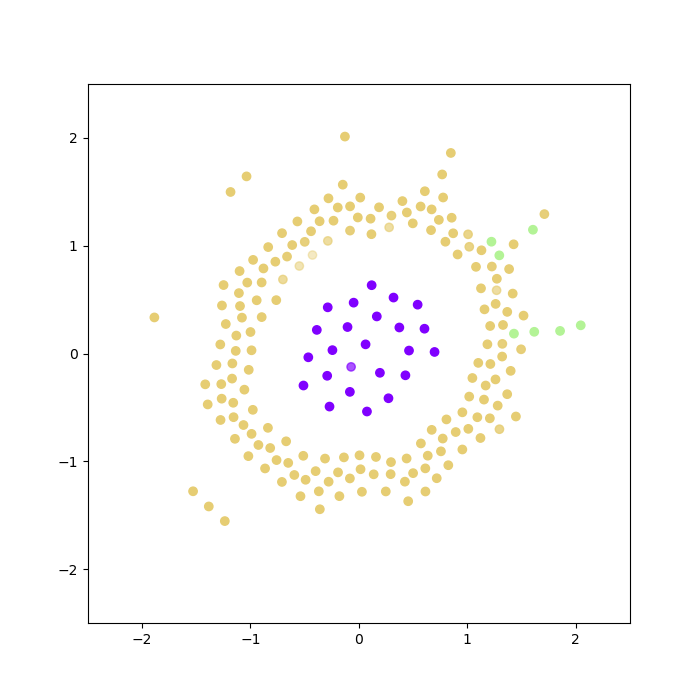}
    \caption{'Flagella' begin to form.}
  \end{subfigure}
  
  \medskip
  
  \begin{subfigure}{0.3\textwidth}
    \centering
    \includegraphics[width=\linewidth]{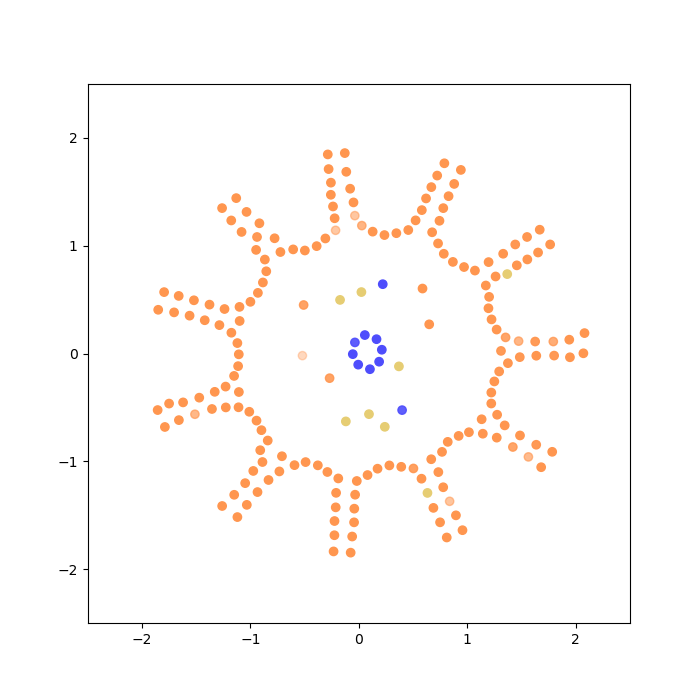}
    \caption{'Flagella' fully formed}
  \end{subfigure}%
  \hfill
  \begin{subfigure}{0.3\textwidth}
    \centering
    \includegraphics[width=\linewidth]{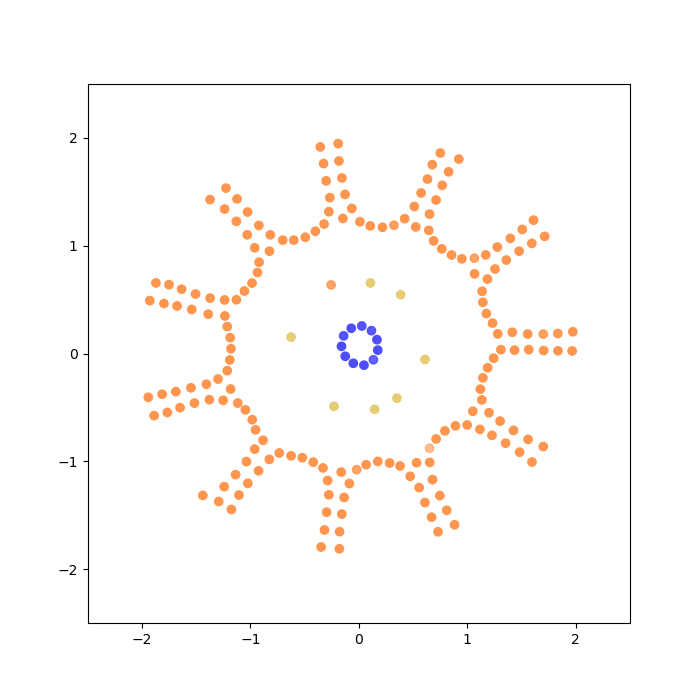}
    \caption{'Organelles' form.}
  \end{subfigure}%
  \hfill
  \begin{subfigure}{0.3\textwidth}
    \centering
    \includegraphics[width=\linewidth]{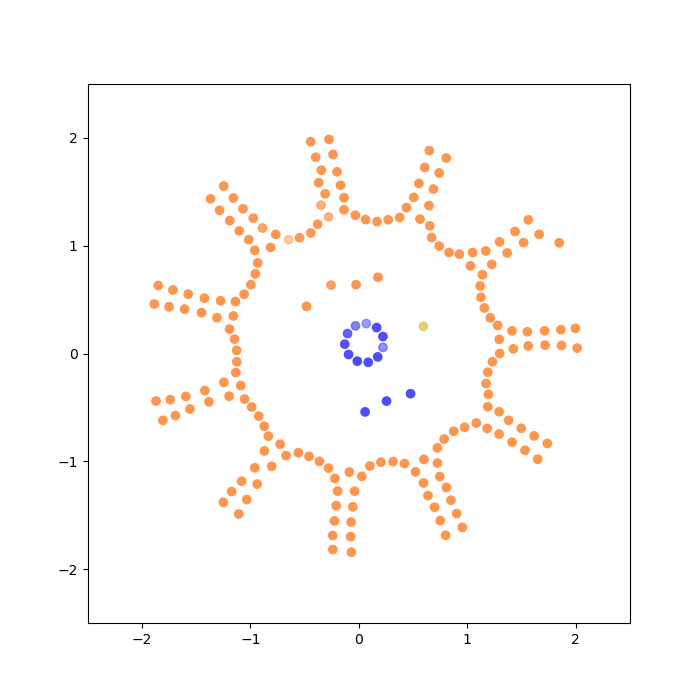}
    \caption{'Organelles' interact.}
  \end{subfigure}
  
  \caption{DMBD fit to initial evolution of particle Lenia 'rotator'.  Time increases left to right.  In the initial disordered state nearly all particles are assigned to a single object.  As time progresses a cell membrane forms which is assigned to a new object type.  The cell membrane then undergoes a phase change as 'flagella'.  This alters the object label of the membrane.  Meanwhile a tightly bound nucleus forms which is also given a unique label.  An additional 8 repulsive particles move about the region between the nucleus and membrane.  This configuration is semi stable with the inner 8 particles transiting around the nucleus.  The membrane pulses and rotates slightly perturbing the 8 inner particles which causes the nucleus to pulsate as well.  This causes the inner 8 particles to regularly shift in their object assignments transitioning from nucleus to organelle to membrane and back.}
  \label{fig:lenia}
\end{figure}

For a final example, we use Particle Lenia to simulate a self-organizing system that exhibit cell-like structure and behavior. We use the ``rotator'' example \cite{chan2019lenia} for the simulation. Individual particles are characterized by different distance-dependent attraction and repulsion functions. This leads to interesting self-organizing behavior depicted in Fig. \ref{fig:lenia}, which shows an initially random soup of particles quickly that forms into a cell-like object, with a ``nucleus'' and simple ``cell membrane.'' After some time, the membrane develops rotating flagella-like structures, and the nucleus tightens into a smaller shape. We use this simulation as a test bed for this multiple-object discovery scheme, in the hopes that it is able to discover that the different ``parts'' of the cell are different objects in a common environment. We assume the presence of 11 different objects, each characterized by 2 dimensional dynamics for the blanket and object variables. Each object is linked to observations via a single role for each of the blanket and object latents, to encourage the discovery of spatially localized objects. In the example shown, the simulation discovers 5 objects, corresponding to (1) a disordered state (green), (2) simple cell membrane (yellow), (3) complex cell membrane (orange), (4) disordered nucleus (purple), and (5) tight nucleus (blue). This illustrates the potential utility of this approach for segmenting complex systems into multiple interacting dynamic subsystems.

\section{Discussion}

In this paper, we situated the FEP, Bayesian mechanics, Markov blankets, and ontological potential functions within the broader set of ideas in Bayesian statistics and information theory, at the intersection of Jaynesian and Bayesian approaches to mathematical physics, and to mathematical modeling more generally. In particular, the work that we presented here shows that, in order to apply the logic and constructs of the FEP to model interesting things, we must step outside of the logic of the FEP itself and call upon an additional principle, by which to pick the best Markov blanket partition, out of the exponentially many different possible ones. Satisfyingly, this turned out to be the same core ideas from which we started in developing the FEP in the first place: Jaynes' principle and surprise minimization. This is \textit{unsurprising} to us, because the dynamic Markov blanket detection algorithm that we developed is itself based on the same underlying ideas that led to the development of the FEP.

We proposed a class of models and associated inference algorithms that treat the problem of dynamic Markov blanket detection as a macroscopic physics discovery problem, which are fundamentally based upon the identification of dynamic boundary elements that result in the simplest macroscopic description of the system as a whole. The output of this process is a set of macroscopic object-type-specific rules that govern the interactions between the boundary of an object and the environment in which it is embedded (mediated by some possibly fictive internal variables). We motivated this approach by arguing that it is the statistics of Markov blankets that define object type, in a manner consistent with systems identification theory and reinforcement learning; and by arguing that, when combined with Jaynes' principle of maximum caliber, this definition leads to familiar descriptions of the associated physical systems in terms of energy functions, Hamiltonians, and Lagrangians. Moreover, the combination of these mathematical tools leads directly to a characterization of objects in terms of an ontological potential function that precisely corresponds to the free energy functional, and therefore, corresponds to an alternative derivation of the FEP that starts from information theory (maximum caliber modeling), as opposed to the traditional approach to deriving the FEP, which begins with the equations of statistical physics.  

On its own, this information theoretic approach to typing objects is incapable of determining which of the many dynamic blankets in a system should be labeled as proper subsystem. To resolve this ambiguity, we appealed to an additional principle: parsimony. That is, good labels should lead to a compact, low-dimensional description of the system as a whole. This is not meant to imply that labeled objects have some kind of metaphysical significance, only that good labels are useful for prediction, generalization, and data compression.

To see how all this might apply in general, consider the humble proton. Despite being composed of a veritable zoo of more fundamental particles, correctly applying the label ``proton`` to the zoo results in a lower entropy of our observations, if for no other reason than the fact that protons have positive charge, a particular mass, etc., and behave accordingly; while a randomly selected particle might have a different mass, charge, etc., and accordingly, behave differently. Indeed, that is why the zoo was given a label: The label has predictive power, in the sense that, given the label and knowledge of the observable properties of a proton (position, momentum, spin, etc.), we can predict how the particle will interact with other things in its environment without having to think about what's going on inside. 

That is, good labels \emph{globally} minimize the conditional entropy of future observations (or surprise). In that sense, in mathematical physics modeling, labels play the explanatory role of testable hypotheses: a good hypothesis makes sense of data, in the sense that the data becomes unsurprising under the hypothesis that the process generating the data can be labeled as being of this or that type. So, labels are useful because they allow us to compactly describe the dynamics or observable behaviors of things, generating less surprise upon observing new data or upon considering past data than if the label had not been used. The dynamics of things depend on their properties, and labels are a useful way of denoting things with similar properties and behavior. Indeed, if the entropy of our observations does not, in fact, go down conditioned on that label, then the mutual information between label and observations is zero by definition, and therefore, the label is meaningless, both pragmatically and in an information theoretic sense. 

Interestingly, conditional entropy is precisely the surprise objective most strongly associated with the FEP. However, in that literature, the role of surprise minimization is treated tautologically, not empirically: the idea is that ``things minimize surprise,'' as opposed to ``labeling something as an object of a specific type minimizes my surprise.'' As a concrete example, consider the humble proton. Correctly applying the label ``proton`` to a particle results in a lower entropy of our observations, if for no other reason than the fact that protons have positive charge, a particular mass, etc., and behave accordingly; while a randomly selected particle might have a different mass, charge, etc., and accordingly, behave differently. Indeed, that is why it was given a label: the label has predictive power, in the sense that, given the label and knowledge of the observable properties of a proton (position, momentum, spin, etc.), we can predict how the particle will interact with other things in its environment. To this end, consider why we bothered to label a proton a ``thing.'' Correctly applied, the label reduces uncertainty/entropy and enhances our ability to predict the behavior of both the proton and the system as a whole. That is, good labels \emph{globally} minimize surprise. This suggests that the \textit{surprise minimization objective} plays a critical empirical role. 

This is distinct from the manner in which the FEP is traditionally discussed. A good tautology provides a good starting point for a definition. The FEP starts with the definition of an object as a blanketed collection of states, whose internal and active states or paths minimizes surprise (as characterized by the free energy of blanket states or paths)---and derives a principle of least action and ensuing Bayesian mechanics. That is, the principle focus of the FEP literature has been on necessary properties of things, and not on empirically discovering which collections we can sensibly label as objects. Conversely and in a complementary fashion, the approach described here starts from a Markov blanket based definition of system types, but takes an \textit{empirical} point of view of an observer or modeler and seeks to provide a justification for the decisions that an observer makes when modeling the dynamics of a system.

\subsection{Response to technical critiques of the FEP}

This work addresses some core technical criticism and limitations of modeling based on the FEP, which up until now were arguably still open.\footnote{Albeit, see section 3.4. ``Some remarks about the state of the art'' in \cite{ramstead2023bayesian} for a discussion of the current state of affairs regarding technical critiques of the FEP, and responses to them).} The Markov blanket based approach to system identification has proven controversial (see, e.g., \cite{bruineberg2022emperors, raja2021markov, dipaolo2022forking, biehl2021technical}). Some have argued that the Markov blanket based definition of object and object type is not as obvious or trivial as its proponents say it is \cite{raja2021markov}.\footnote{Interestingly to us, when considering what possible alternative there could be to the FEP from an information theoretic perspective, the authors \cite{raja2021markov} land upon \textit{maximum entropy} as a possible alternative. This choice of candidate alternative approach is telling, especially in light of recent developments on the duality of the FEP and maximum entropy/caliber. Indeed, there seems to be no other game in town.} Another line of criticism argues that actually \textit{identifying} Markov blankets (both mathematically and empirically, in real world data) rests upon nontrivial modeling decisions, and that, as a result, the Markov blanket formalism is less easily or generally applicable than claimed \cite{bruineberg2022emperors}. Still others have noted that demonstrations of the near universal applicability of the FEP seems at odds with the assumed form of the Markov blanket and nonequilibrium steady state condition that together guarantee a partition between organism and environment or between internal and external states \cite{buckleyparticular, biehl2021technical, dipaolo2022forking, nave2025drive}. Indeed, the Markov blanket based definition of a macroscopic object has been criticized as being ill posed (since many interesting types of systems will appear to have \textit{many} Markov blankets \cite{bruineberg2022emperors}) or inapplicable to systems that are strongly coupled to their environment, highly variable, or exchange matter with their environment \cite{raja2021markov, dipaolo2022forking, nave2025drive}. Plainly, the systems that are of most interest to us are open systems that exchange matter and energy with their environments, exist far from thermodynamic equilibrium, and usually have mobile and dynamic---and possibly non contiguous---boundaries. Candidate counterexamples to the applicability of the Markov blanket based approach turned out to be things with mobile or wandering boundaries, like flames and organisms. It was especially problematic because the FEP was developed originally to model the dynamics of the brain and the behavior of living things, and is probably best known in this application. 

As one would hope, these criticisms have led to substantial debate and an explicit acknowledgment that the FEP, as formulated for systems at steady state, applies only to a rather ``particular'' class of macroscopic objects that only sparsely or weakly interact with their environment \cite{fristonparticular, buckleyparticular}. Here, we argue that many of these criticism ultimately stem from the FEP literature's prior focus on ``static'' Markov blanket construct. From a mathematical point of view, several of these objections were resolved by moving from a state-wise formulation to a path-based or path integral formulation of the FEP \cite{sakthivadivel2022regarding, friston2023path}, allowing us to avoid having to make assumptions about the steady state statistics of a system (see \cite{ramstead2023bayesian} for a detailed discussion) or rely on approximate Markov blanket structure in the stationary statistics of the system as a whole. In the path-based formulations of the FEP, quantities of interest and the equations that relate them now are defined for paths of a system, i.e., they are defined in objects that make up a structured space or set of trajectories that each represent a specific way the system might evolve over time. Moreover, a consideration of dynamic blankets and maximum caliber modeling \cite{ramstead2023bayesian, sakthivadivel2022worked} allows us to identify sensible objects and boundaries in complex systems that include phase transitions, and to models of objects with transient and moving boundaries. Our contribution here is to provide a mathematical framework and numerical demonstrations that show definitively that such dynamic objects can be modeled within this framework, thereby resolving this debate in the literature empirically. 

\subsection{Niche construction and the role of the environment in active inference}

In our simulation of a burning fuse, the environment played an unexpected role: we note that the position of the flame is best tracked by the state of the environment, rather than by internal states of the combustion network. We also noted above that, because the equations that govern the dynamics of Markov blankets are symmetric with respect to internal and external states, the blanket encodes the policies of both the internal and the external subsystems, concluding that the Markov blanket based definition of an object type is always environment-specific. This result sits well with recent work reconsidering the role of the environment in active inference and with work on active inference in sociocultural systems \cite{veissiere2020thinking}. This work considers the multi scale recursively nested structure of the dynamics that coupled individual agents, the sociocultural systems they form, and the ecological niche that these shape through action \cite{constant2018variational, ramstead2019variational, veissiere2020thinking, parr2020choosing}. There is no \textit{a priori} reason to always center our model on agents for every kind of task and situation. Instead, we can model agents as special parts of the wider environment---that are highly salient to other agents; see e.g., \cite{esaki2024environment}. This also speaks to previous work in this tradition on niche construction work, emphasizing that the property of the synchronization between objects that the FEP describes---i.e., that it is \textit{symmetric}---can be exploited by agents. This work models the passive (habitual, unintentional) and active  construction (e.f., deliberate design) of an ecological niche by its denizens, such that certain kinds of patterned behavior are solicited, as opposed to others that are discouraged \cite{constant2018variational, ramstead2016cultural}.

\subsection{Future directions}

Our specific realization of a member of this class of Markov blanket detection algorithms relied on linear approximations and a decoupling of blanket (label) dynamics and macroscopic dynamics. This choice resulted in an algorithm that sensibly partition systems, but that cannot be relied upon for prediction. This is for two reasons: (1) the assumption of linear dynamics causes the effects of any non-linearities in the system to be attributed to noise, causing an effective enhancement of diffusive strength; and (2) the assumption that the boundary dynamics are decoupled from the macroscopic dynamics means that latent assignment variables quickly diffuse to a uniform stationary distribution in the absence of observed data. We plan to address these issues in future work by imposing Markov blanket structure on Bayesian instantiations of switching linear dynamical systems models. This work is also related to work on mathematically modeling downward causality and emergent phenomena via dimensionality reduction \cite{barnett2023dynamical, rosas2020reconciling}, in ways that we plan to explore.

\printbibliography

@article{biehl2021technical,
  title={A Technical Critique of Some Parts of the Free Energy Principle},
  author={Biehl, Martin and Pollock, Felix A. and Kanai, Ryota},
  journal={Entropy},
  volume={23},
  number={3},
  pages={293},
  year={2021},
  publisher={Multidisciplinary Digital Publishing Institute},
  doi={10.3390/e23030293},
  pmcid={PMC7997279},
  pmid={33673663},
  editor={Knuth, Kevin H.}
}

@article{raissi2019pinn,
  title={Physics-informed neural networks: A deep learning framework for solving forward and inverse problems involving nonlinear partial differential equations},
  author={Raissi, Maziar and Perdikaris, Paris and Karniadakis, George E},
  journal={Journal of Computational Physics},
  volume={378},
  pages={686--707},
  year={2019},
  publisher={Elsevier}
}

@article{gonzalez2016,
  title={Jarzynski equality in the context of maximum path entropy},
  author={Diego González and Sergio Davis},
  journal={arXiv preprint arXiv:1607.07287},
  year={2016},
  url={https://arxiv.org/abs/1607.07287}
}

@InProceedings{linderman17rslds,
  title = {{Bayesian Learning and Inference in Recurrent Switching Linear Dynamical Systems}},
  author = {Linderman, Scott and Johnson, Matthew and Miller, Andrew and Adams, Ryan and Blei, David and Paninski, Liam},
  booktitle = {Proceedings of the 20th International Conference on Artificial Intelligence and Statistics},
  pages = {914--922},
  year = {2017},
  editor = {Singh, Aarti and Zhu, Jerry},
  volume = {54},
  series = {Proceedings of Machine Learning Research},
  month = {4},
  publisher = {PMLR},
  url = {https://proceedings.mlr.press/v54/linderman17a.html}
}

@article{ramstead2023bayesian,
  author = {Ramstead, Maxwell J. D. and Sakthivadivel, Dalton A. R. and Heins, Conor and Koudahl, Magnus and Millidge, Beren and Da Costa, Lancelot and Klein, Brennan and Friston, Karl J.},
  title = {On Bayesian mechanics: a physics of and by beliefs},
  journal = {Interface Focus},
  year = {2023},
  day = {14},
  volume = {13},
  number = {3},
  pages = {20220029},
  doi = {10.1098/rsfs.2022.0029},
  url = {https://doi.org/10.1098/rsfs.2022.0029}
}

@article{beck2003nonlinear,
  title = "Nonlinear dynamics in a simple model of solid flame microstructure",
  author = "Beck, J.M. and Volpert, V.A.",
  journal = "Physica D: Nonlinear Phenomena",
  volume = "182",
  number = "1--2",
  pages = "86--102",
  year = "2003",
  doi = "10.1016/S0167-2789(03)00189-5",
}

@article{chan2019lenia,
  title={Lenia: Biology of Artificial Life},
  author={Chan, Bert Wang-Chak},
  journal={Complex Systems},
  volume={28},
  number={3},
  year={2019},
  pages={251--286},
  publisher={Complex Systems Publications},
  address={Hong Kong},
  doi={10.25088/ComplexSystems.28.3.251},
  url={https://doi.org/10.25088/ComplexSystems.28.3.251}
}

@article{sakthivadivel2022geometry,
  title={Towards a Geometry and Analysis for Bayesian Mechanics},
  author={Sakthivadivel, Dalton A R},
  journal={arXiv preprint arXiv:2204.11900},
  year={2022},
  url={https://arxiv.org/abs/2204.11900},
  archivePrefix={arXiv},
  primaryClass={math-ph},
  eprint={2204.11900},
  doi={10.48550/arXiv.2204.11900},
}

@book{pearl1998,
  title = "Quantified Representation of Uncertainty and Imprecision",
  author = "Pearl, Judea",
  publisher = "Springer",
  year = "1998",
  series = "Graphical models for probabilistic and causal reasoning",
  pages = "367--389",
}

@article{hoffman2013stochastic,
  author = {Hoffman, Matthew D. and Blei, David M. and Wang, Chong and Paisley, John},
  title = {Stochastic Variational Inference},
  journal = {Journal of Machine Learning Research},
  volume = {14},
  year = {2013},
  pages = {1303--1347},
  url = {http://www.jmlr.org/papers/volume14/hoffman13a/hoffman13a.pdf}
}

@article{wang2020towards,
  author = {Wang, Hao and Ling, Zhaolong and Yu, Kui and Wu, Xindong},
  title = {Towards efficient and effective discovery of Markov blankets for feature selection},
  journal = {Information Sciences},
  volume = {509},
  year = {2020},
  pages = {227--242},
  doi = {10.1016/j.ins.2019.09.053},
  url = {https://doi.org/10.1016/j.ins.2019.09.053}
}

@article{glymour2019review,
  author = {Glymour, Clark and Zhang, Kun and Spirtes, Peter},
  title = {Review of Causal Discovery Methods Based on Graphical Models},
  journal = {Frontiers in Genetics},
  volume = {10},
  year = {2019},
  pages = {524},
  doi = {10.3389/fgene.2019.00524},
  url = {https://doi.org/10.3389/fgene.2019.00524},
  note = {This article is part of the Research Topic Shift the Current Paradigm of Genetic Studies of Complex Diseases from Association to Causation},
  issn = {1664-8021},
  publisher = {Frontiers Media},
  number = {June},
  section = {Statistical Genetics and Methodology}
}

@article{jaynes1957,
  title={Information theory and statistical mechanics},
  author={Jaynes, Edwin T.},
  journal={Physical Review},
  volume={106},
  number={4},
  pages={620},
  year={1957},
  publisher={American Physical Society},
  doi={10.1103/PhysRev.106.620}
}

@article{jones2024bong,
  title     = {Bayesian Online Natural Gradient BONG},
  author    = {Matt Jones and Peter Chang and Kevin Murphy},
  journal   = {arXiv preprint arXiv:2405.19681},
  year      = {2024},
  url       = {https://arxiv.org/abs/2405.19681}
}

@article{davis2015hamiltonian,
  author = {Davis, Sergio and González, Diego},
  title = {Hamiltonian formalism and path entropy maximization},
  journal = {Journal of Physics A: Mathematical and Theoretical},
  volume = {48},
  number = {42},
  pages = {42500},
  year = {2015},
  publisher = {IOP Publishing Ltd},
  doi = {10.1088/1751-8113/48/42/42500},
}

@article{snoswell2021revisiting,
  title={Revisiting Maximum Entropy Inverse Reinforcement Learning: New Perspectives and Algorithms},
  author={Snoswell, Aaron J. and Singh, Surya P. N. and Ye, Nan},
  journal={arXiv preprint arXiv:2108.10056},
  year={2021}
}

@inproceedings{ziebart2008maximum,
  title={Maximum entropy inverse reinforcement learning},
  author={Ziebart, Brian D and Maas, Andrew and Bagnell, J Andrew and Dey, Anind K},
  booktitle={Proceedings of the 23rd AAAI Conference},
  year={2008}
}

@article{eslami2016attend,
  title={Attend, Infer, Repeat: Fast Scene Understanding with Generative Models},
  author={Eslami, S. M. Ali and Heess, Nicolas and Weber, Theophane and Tassa, Yuval and Szepesvari, David and Kavukcuoglu, Koray and Hinton, Geoffrey E.},
  journal={arXiv preprint arXiv:1603.08575},
  year={2016},
  eprint={1603.08575},
  archivePrefix={arXiv},
  primaryClass={cs.CV},
  url={https://doi.org/10.48550/arXiv.1603.08575}
}

@article{bruineberg2022emperors,
  title={The Emperor's New Markov Blankets},
  author={Bruineberg, Johan and Dołęga, Karolina and Dewhurst, Joe and Baltieri, Mauro},
  journal={Behavioral and Brain Sciences},
  volume={45},
  pages={E183},
  year={2022},
  publisher={Cambridge University Press},
  doi={10.1017/S0140525X21002351}
}

@article{buckleyparticular,
  title={How particular is the physics of the free energy principle?},
  author={Aguilera, Miguel and Millidge, Beren and Tschantz, Alexander and Buckley, Christopher L.},
  journal={Physics of Life Reviews},
  volume={40},
  pages={24--50},
  year={2022},
  publisher={Publisher}
}

@article{fristonparticular,
  title={A free energy principle for a particular physics},
  author={Friston, Karl},
  journal={arXiv preprint arXiv:1906.10184},
  year={2019},
  eprint={1906.10184},
  archivePrefix={arXiv},
}

@inproceedings{Todorov2010InverseOC,
  title={Inverse Optimal Control with Linearly-Solvable MDPs},
  author={Krishnamurthy Dvijotham and Emanuel Todorov},
  booktitle={International Conference on Machine Learning},
  year={2010}
}

@article{kappen2012optimal,
  title = {Optimal control as a graphical model inference problem},
  author = {Kappen, B. and Gomez, V. and Opper, M.},
  journal = {Machine Learning Journal},
  year = {2012},
  volume = {},
  number = {},
  pages = {},
  eprint = {arXiv:0901.0633 [math.OC]},
  archivePrefix = {arXiv},
  primaryClass = {math.OC},
  secondaryClass = {eess.SY},
  comment = {26 pages, 12 Figures},
  subjects = {Optimization and Control (math.OC); Systems and Control (eess.SY)},
  acmClasses = {F.1.2; G.3; I.2.8},
}

@article{niven2010,
  title={Minimization of a free-energy-like potential for non-equilibrium flow systems at steady state},
  author={Niven, Robert K.},
  journal={Philosophical Transactions of the Royal Society B: Biological Sciences},
  volume={365},
  number={1545},
  pages={1323--1331},
  year={2010},
  doi={10.1098/rstb.2009.0296},
  PMID={20368250},
  PMCID={PMC2871899}
}

@article{ramstead2024approach,
  title={An approach to non-equilibrium statistical physics using variational Bayesian inference},
  author={Ramstead, Maxwell JD and Sakthivadivel, Dalton AR and Friston, Karl J},
  journal={arXiv preprint arXiv:2406.11630},
  year={2024}
}

@article{barnett2023dynamical,
  title={Dynamical independence: discovering emergent macroscopic processes in complex dynamical systems},
  author={Barnett, Lionel and Seth, Anil K},
  journal={Physical Review E},
  volume={108},
  number={1},
  pages={014304},
  year={2023},
  publisher={APS}
}

@article{friston2023path,
  title={Path integrals, particular kinds, and strange things},
  author={Friston, Karl and Da Costa, Lancelot and Sakthivadivel, Dalton AR and Heins, Conor and Pavliotis, Grigorios A and Ramstead, Maxwell and Parr, Thomas},
  journal={Physics of Life Reviews},
  year={2023},
  publisher={Elsevier}
}

@inproceedings{sakthivadivel2022worked,
  title={A worked example of the Bayesian mechanics of classical objects},
  author={Sakthivadivel, Dalton AR},
  booktitle={International Workshop on Active Inference},
  pages={298--318},
  year={2022},
  organization={Springer}
}

@article{dacosta2021bayesian,
  title={Bayesian mechanics for stationary processes},
  author={Da Costa, Lancelot and Friston, Karl and Heins, Conor and Pavliotis, Grigorios A},
  journal={Proceedings of the Royal Society A},
  volume={477},
  number={2256},
  pages={20210518},
  year={2021},
  publisher={The Royal Society}
}

@article{raja2021markov,
  title={The Markov blanket trick: On the scope of the free energy principle and active inference},
  author={Raja, Vicente and Valluri, Dinesh and Baggs, Edward and Chemero, Anthony and Anderson, Michael L},
  journal={Physics of Life Reviews},
  volume={39},
  pages={49--72},
  year={2021},
  publisher={Elsevier}
}

@article{sakthivadivel2022weak,
  title={Weak Markov blankets in high-dimensional, sparsely-coupled random dynamical systems},
  author={Sakthivadivel, Dalton AR},
  journal={arXiv preprint arXiv:2207.07620},
  year={2022}
}

@article{heins2022sparse,
  title={Sparse coupling and Markov blankets: A comment on" How particular is the physics of the Free Energy Principle?" by Aguilera, Millidge, Tschantz and Buckley},
  author={Heins, Conor and Da Costa, Lancelot},
  journal={arXiv preprint arXiv:2205.10190},
  year={2022}
}

@article{dipaolo2022forking,
  title={Laying down a forking path: Tensions between enaction and the free energy principle},
  author={Di Paolo, Ezequiel and Thompson, Evan and Beer, Randall},
  journal={Philosophy and the Mind Sciences},
  volume={3},
  year={2022}
}

@article{sakthivadivel2022regarding,
  title={Regarding Flows Under the Free Energy Principle: A Comment on" How Particular is the Physics of the Free Energy Principle?" by Aguilera, Millidge, Tschantz, and Buckley},
  author={Sakthivadivel, Dalton AR},
  journal={arXiv preprint arXiv:2205.07793},
  year={2022}
}

@article{constant2018variational,
  title={A variational approach to niche construction},
  author={Constant, Axel and Ramstead, Maxwell JD and Veissiere, Samuel PL and Campbell, John O and Friston, Karl J},
  journal={Journal of the Royal Society Interface},
  volume={15},
  number={141},
  pages={20170685},
  year={2018},
  publisher={The Royal Society}
}

@article{esaki2024environment,
  title={Environment-Centric Active Inference},
  author={Esaki, Kanako and Matsumura, Tadayuki and Kato, Takeshi and Minusa, Shunsuke and Shao, Yang and Mizuno, Hiroyuki},
  journal={arXiv preprint arXiv:2408.12777},
  year={2024}
}

@article{veissiere2020thinking,
  title={Thinking through other minds: A variational approach to cognition and culture},
  author={Veissi{\`e}re, Samuel PL and Constant, Axel and Ramstead, Maxwell JD and Friston, Karl J and Kirmayer, Laurence J},
  journal={Behavioral and brain sciences},
  volume={43},
  pages={e90},
  year={2020},
  publisher={Cambridge University Press}
}

@article{parr2020choosing,
  title={Choosing a Markov blanket.},
  author={Parr, Thomas},
  journal={Behavioral \& Brain Sciences},
  volume={43},
  year={2020}
}

@article{ramstead2016cultural,
  title={Cultural affordances: Scaffolding local worlds through shared intentionality and regimes of attention},
  author={Ramstead, Maxwell JD and Veissi{\`e}re, Samuel PL and Kirmayer, Laurence J},
  journal={Frontiers in psychology},
  volume={7},
  pages={1090},
  year={2016},
  publisher={Frontiers Media SA}
}

@article{ramstead2019variational,
  title={Variational ecology and the physics of sentient systems},
  author={Ramstead, Maxwell JD and Constant, Axel and Badcock, Paul B and Friston, Karl J},
  journal={Physics of life Reviews},
  volume={31},
  pages={188--205},
  year={2019},
  publisher={Elsevier}
}

@book{beal2003variational,
  title={Variational algorithms for approximate Bayesian inference},
  author={Beal, Matthew James},
  year={2003},
  publisher={University of London, University College London (United Kingdom)}
}

@article{koopman1931hamiltonian,
  title={Hamiltonian systems and transformation in Hilbert space},
  author={Koopman, Bernard O},
  journal={Proceedings of the National Academy of Sciences},
  volume={17},
  number={5},
  pages={315--318},
  year={1931},
  publisher={National Acad Sciences}
}

@article{vaswani2017attention,
  title={Attention is all you need},
  author={Vaswani, A},
  journal={Advances in Neural Information Processing Systems},
  year={2017}
}

@article{friston2021parcels,
  title={Parcels and particles: Markov blankets in the brain},
  author={Friston, Karl J and Fagerholm, Erik D and Zarghami, Tahereh S and Parr, Thomas and Hip{\'o}lito, In{\^e}s and Magrou, Lo{\"\i}c and Razi, Adeel},
  journal={Network Neuroscience},
  volume={5},
  number={1},
  pages={211--251},
  year={2021},
  publisher={MIT Press One Rogers Street, Cambridge, MA 02142-1209, USA journals-info~…}
}

@article{jaynes1980minimum,
  title={The minimum entropy production principle},
  author={Jaynes, Edwin T},
  journal={Annual Review of Physical Chemistry},
  volume={31},
  number={1},
  pages={579--601},
  year={1980},
  publisher={Springer}
}

@inproceedings{jarzynski2012equalities,
  title={Equalities and inequalities: Irreversibility and the second law of thermodynamics at the nanoscale},
  author={Jarzynski, Christopher},
  booktitle={Time: Poincar{\'e} Seminar 2010},
  pages={145--172},
  year={2012},
  organization={Springer}
}

@article{kiefer2020psychophysical,
  title={Psychophysical identity and free energy},
  author={Kiefer, Alex B},
  journal={Journal of the Royal Society Interface},
  volume={17},
  number={169},
  pages={20200370},
  year={2020},
  publisher={The Royal Society}
}

@article{rosas2020reconciling,
  title={Reconciling emergences: An information-theoretic approach to identify causal emergence in multivariate data},
  author={Rosas, Fernando E and Mediano, Pedro AM and Jensen, Henrik J and Seth, Anil K and Barrett, Adam B and Carhart-Harris, Robin L and Bor, Daniel},
  journal={PLoS computational biology},
  volume={16},
  number={12},
  pages={e1008289},
  year={2020},
  publisher={Public Library of Science San Francisco, CA USA}
}

@book{nave2025drive,
  title={A drive to survive: The free energy principle and the meaning of life},
  author={Nave, Kathryn},
  year={2025},
  publisher={MIT Press}
}

\end{document}